\documentclass[prb,a4paper,twocolumn,showpacs,floatfix]{revtex4-1}
\usepackage[utf8]{inputenc}
\usepackage[T1]{fontenc}
\usepackage{lmodern}
\usepackage{amssymb}
\usepackage{amsmath}
\usepackage{amsbsy}
\usepackage{bm}
\usepackage{esint}
\usepackage{graphicx}
\newcommand{\I}{\mathrm{i}}
\newcommand{\E}{\mathrm{e}}
\begin{document}
\title{Enhanced spin Hall effect in strong magnetic disorder}
\author{T. L. van den Berg}
\author{L. Raymond}
\author{A. Verga}\email{Alberto.Verga@univ-amu.fr}
\affiliation{Université d'Aix-Marseille, IM2NP-CNRS, Campus St. Jérôme, Case 142, 13397 Marseille, France}
\date{\today}
\begin{abstract}
We consider a two-dimensional electron gas in an inversion asymmetric layer and in the presence of spatially distributed magnetic impurities. We investigate the relationship between the geometrical properties of the wave-function and the system's spin dependent transport properties. A localization transition, arising when disorder is increased, is exhibited by the appearance of a fractal state with finite inverse participation ratio. Below the transition, interference effects modify the carrier's diffusion, as revealed by the dependence on the scattering time of the power law exponents characterizing the spreading of a wave packet. Above the transition, in the strong disorder regime, we find that the states are spin polarized and localized around the impurities. A significant enhancement of the spin current develops in this regime.
\end{abstract}
\pacs{72.25.Dc, 72.25.Rb, 75.50.Pp}
\maketitle

%
\section{Introduction}
\label{S-intro}
The intrinsic spin Hall effect refers to the generation of a dissipationless spin current in response to an external transverse electric field, it was predicted to occur in p-type semiconductors\cite{Murakami-2003pb} and n-type heterostructures,\cite{Sinova-2004fb} and it was observed experimentally in non-magnetic two-dimensional systems.\cite{Kato-2004fu,Wunderlich-2005dq,Brune-2010uq} The effect originates in the correlation between the particle's motion, driven by the external electric field, and its spin, through the spin-orbit interaction. In the case of a two-dimensional electron gas with spin-orbit Rashba coupling (in an inversion asymmetric layer), the spin Hall conductivity is suppressed by impurity scattering.\cite{Inoue-2004fk} However, breaking time invariance by the presence of magnetic impurities, restores the spin Hall effect.\cite{Gorini-2008lq,Gorini-2008kx,van-den-Berg-2011fk} Remarkably, spin transport in a magnetic disordered layer is reinforced by multiple scattering interactions, and, for a large range of impurity concentrations (in the weak disorder regime), the spin conductivity remains near its universal clean value, \(\sigma_{sH}=-e/8\pi\) (\(-e\) is the electron charge).\cite{van-den-Berg-2011fk} Two questions arise naturally concerning the behavior of the spin Hall conductivity: first, how does it evolve in the strong disorder regime, in particular, does the system undergo a localization transition; second, what is the influence of the carrier mediated interactions between impurities, that may induce a ferromagnetic transition. In this paper we shall focus on the first problem, the dependence of \(\sigma_{sH}\) on the concentration and strength of the paramagnetic disorder.

The understanding of spin transport mechanisms in semiconductors, in which the existence of a strong spin-orbit coupling enables the electrical manipulation of the spin, is important for spintronics applications.\cite{dietl2008spintronics,Zutic-2004rz,Chiba-2008uq,Awschalom-2009uq,Wu-2010fk} In particular, recent experiments in magnetically doped semiconductors,\cite{MacDonald-2005fk,Jungwirth-2006ca} reveal interesting physical phenomena, such as the generation of spin currents,\cite{Ganichev-2009vn} the fractal geometry of localized states,\cite{Richardella-2010kl} and intriguing properties of the intrinsic anomalous Hall effect.\cite{Mihaly-2008fk,Liu-2011kx} These experiments show the interplay between confinement, spin-orbit splitting, magnetic disorder and carrier mediated interactions. Their combined action drives spin transport, determines the magnetization structure, and governs the localization transition, that modifies in turn the magnetic properties and the spin conductivity. A minimal model Hamiltonian accounting for these effects, must contain a hopping term, a Rashba term coupling the carrier's momentum to their spin, and an exchange term with the interaction of the itinerant spins with the magnetic moments of randomly distributed impurities.\cite{Bychkov-1984uq,Dugaev-2005ys,Inoue-2006fk,Onoda-2008fk}

Most studies on dilute magnetic semiconductor quantum wells, deal with the influence of the carrier induced ferromagnetic transition and phase separation on the transport mechanisms, and more specifically, with the existence of carrier localized states and their impact on the anomalous Hall effect.\cite{Nagaosa-2010vn} As a natural extension to these investigations, we focus in this paper on the intrinsic spin Hall current, which arises, as the analogous intrinsic anomalous Hall conductivity, from the Berry phase contribution to the velocity operator.\cite{Murakami-2006df} Our goal is to identify the Anderson localization of the carriers as a function of the disorder strength, and to relate the geometry of the quantum states to the spin transport properties beyond the weak disorder regime.

We start with a one band tight binding model taking into account spin-orbit coupling and impurities interactions. In this framework we study the spreading of a wave packet by solving the time-dependent Schrödinger equation. The knowledge of the time evolution allows the computation of the typical diffusion exponents, the fractal correlation dimension and the inverse participation ratio that characterize, at long time scales, the transport and localization of the quantum states. We find that the exponents vary from  near ballistic values in the weak disorder regime, to a diffusion limit at stronger disorder. Increasing the disorder, one observes that the wave function becomes fractal and that the inverse participation ratio remains finite, indicating the existence of a localization transition. This is further confirmed by the local density of states that concentrates around the impurities.

The presence of interference effects in the spreading of the wave function are investigated, in the weak disorder approximation, using the linear response theory. We compute the quantum corrections to the spin conductivity, and find that a weak antilocalization effect arises, which is essentially proportional to the spin-orbit induced spin splitting.

The spin Hall conductivity, which initially decreases with the disorder strength, undergoes a drastic increase in the strong disorder limit, especially when impurity bands appear in the density of states. This enhancement of the spin conductivity in a regime where the carrier states are localized is attributed to the tight correlation between the carrier and impurity spins, that locally breaks the symmetry between spin up and down states.

%
\section{Model}
\label{S-model}
%
%
We consider an electron conduction band of a semiconductor heterostructure, doped with a concentration \(c\) of magnetic impurities, in the tight binding approximation. We assume a simple geometry, where electrons and impurities reside in a square lattice of \(N\) sites, size \(L\times L\) and spacing \(a\). The model hamiltonian, 
\begin{equation}
H = H_e + H_{so} + H_{xc}
\label{e-H}
\end{equation}
contains a kinetic, 
\[
H_e = - t_{hop} \sum_{\langle i,j \rangle} c_i^{\dagger} c_j
\]
a spin-orbit,
\[
H_{so} = \frac{\mathrm{i} \lambda}{a}  \sum_i 
   \left( c_i^{\dagger} \sigma_x c_{i+\hat{y}} 
   - c_i^{\dagger} \sigma_y c_{i+\hat{x}} \right)
    + \mathrm{h.c.}
\]
and a double exchange term, 
\[ 
H_{xc} = \frac{J_s}{\hbar}  \sum_{i \in \mathcal{I}} \bm{n}_i \cdot  c^{\dagger}_i \bm{\sigma}c_i, 
\]
where \(\bm{\sigma}\) is the vector of Pauli matrices, \(c_i^{\dagger}=(c^{\dagger}_{i\uparrow}\; c^{\dagger}_{i\downarrow})\) (\(c_i\)) is the creation (annihilation) operator on site \(i\) and spin up \(\uparrow\), or down \(\downarrow\); \(t_{hop}\) is the hopping energy between neighboring sites \(i,j\), \(\lambda\) is the Rashba spin-orbit coupling constant, and \(J_s\) the exchange interaction energy constant. The spin-orbit term, proportional to \(\lambda\), contains hops combined with a spin-flip to neighbors in the \(\hat x\) and \(\hat y\) directions. The magnetic moments are randomly distributed over sites \(i\) belonging to the set \(\mathcal{I}\) of impurities, and their orientation \(\bm{n}_i\), is uniformly distributed on the unit sphere. We choose units such that \(\hbar = t_{hop} = a = 1\).

The above Hamiltonian, can be considered as a minimal model of a dilute magnetic semiconductor electron gas, confined in a well between two asymmetric layers, in the paramagnetic state.\cite{Gui-2004fk} It is well suited for investigating the anomalous Hall effect, and the intrinsic spin Hall effect in the presence of magnetic impurities.\cite{Inoue-2009fk}

The presence of spin-orbit coupling and magnetic impurities breaks spin rotation and time reversal symmetries, allowing the occurrence in two dimensions, of the Anderson metal-insulator transition.\cite{Anderson-1958pb,Bergmann-1984vn,Evers-2008nx} To characterize the localization of the quantum states and its influence on spin transport properties, it is convenient to investigate both dynamical and spectral properties.\cite{Chalker-1988mz,Ketzmerick-1992fk,Brandes-1996ly} We therefore investigated the spreading of a wave packet and the spatial distribution of the density of states. From the time evolution of the wave function, we can extract information on the diffusion and the time autocorrelation exponents, which are related with the  spectrum of fractal dimensions.\cite{Ketzmerick-1997fk,Huckestein-1999zr} To distinguish between extended and localized states we use the inverse participation ratio.\cite{Thouless-1974kx,Wegner-1980kx,Janssen-1994fk} The localization transition is usually associated with a multifractal structure of the wave functions, that can deeply influence the transport mechanisms.\cite{Aoki-1986vn,Castellani-1986uq,mirlin2010anderson} We also measure the local density of states to characterize the spatial distribution of the eigenstates over the energy spectrum.\cite{Mirlin-2000uq,Schubert-2010kx}

\begin{figure*}
\centering
\includegraphics[width=0.165\textwidth]{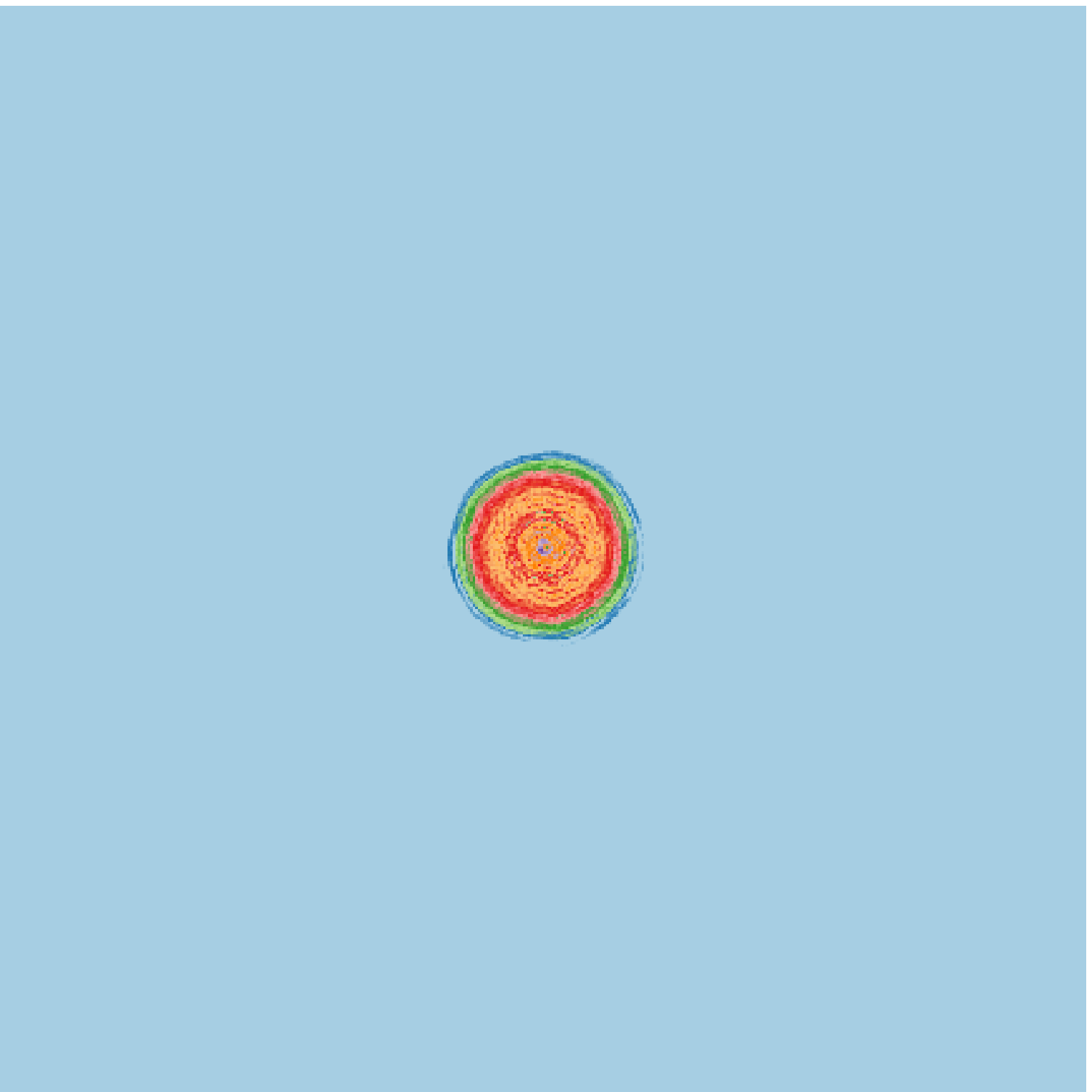}%
\includegraphics[width=0.165\textwidth]{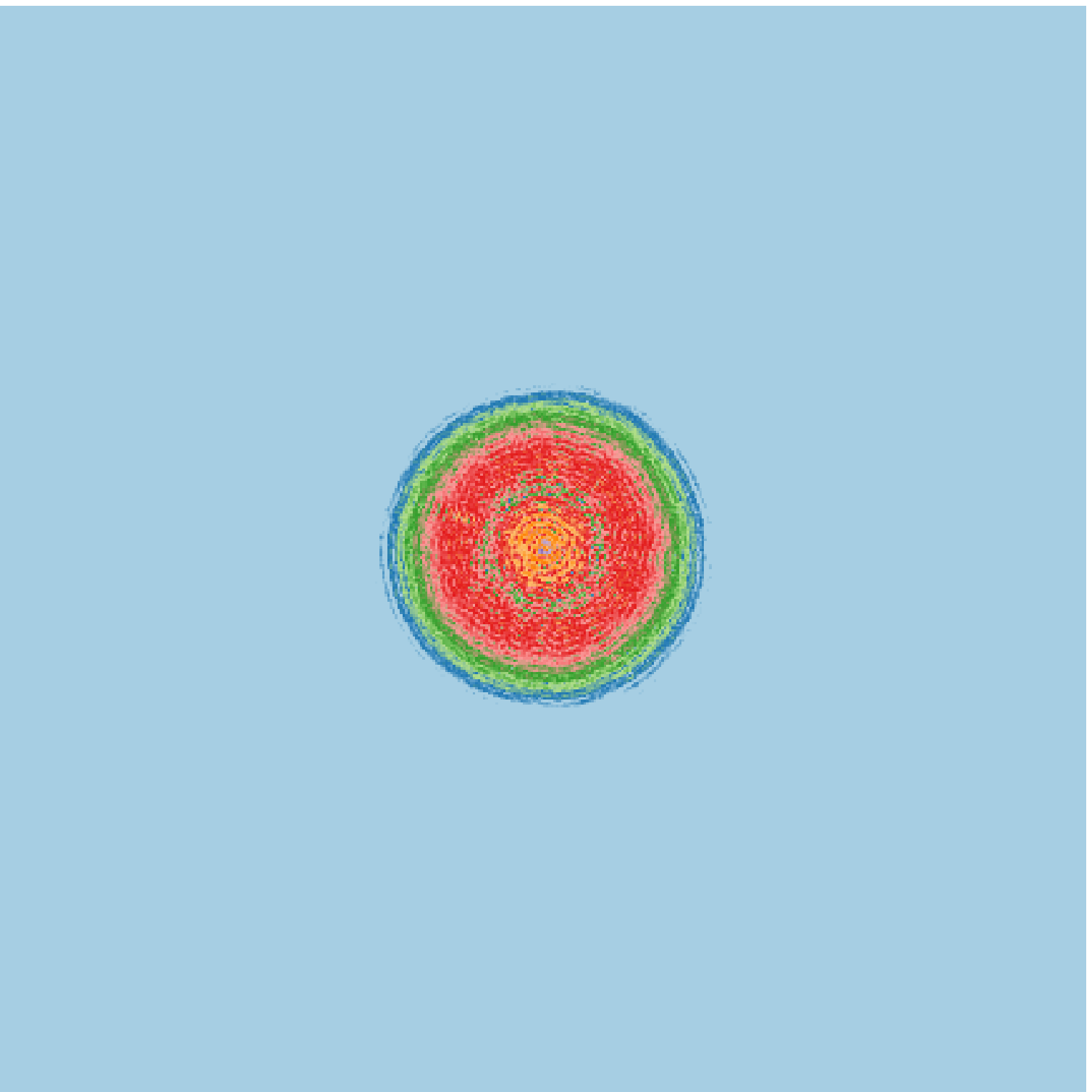}%
\includegraphics[width=0.165\textwidth]{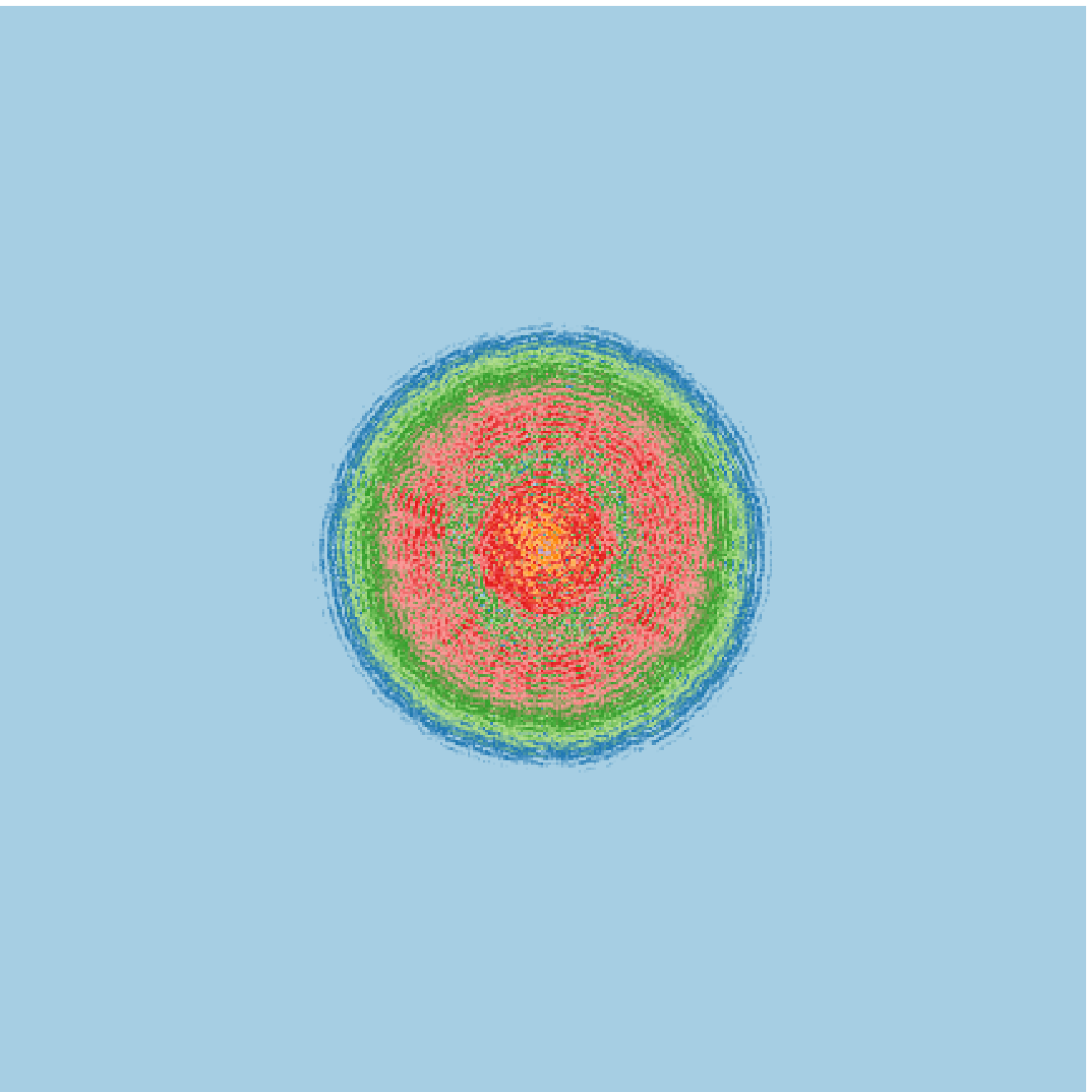}%
\includegraphics[width=0.165\textwidth]{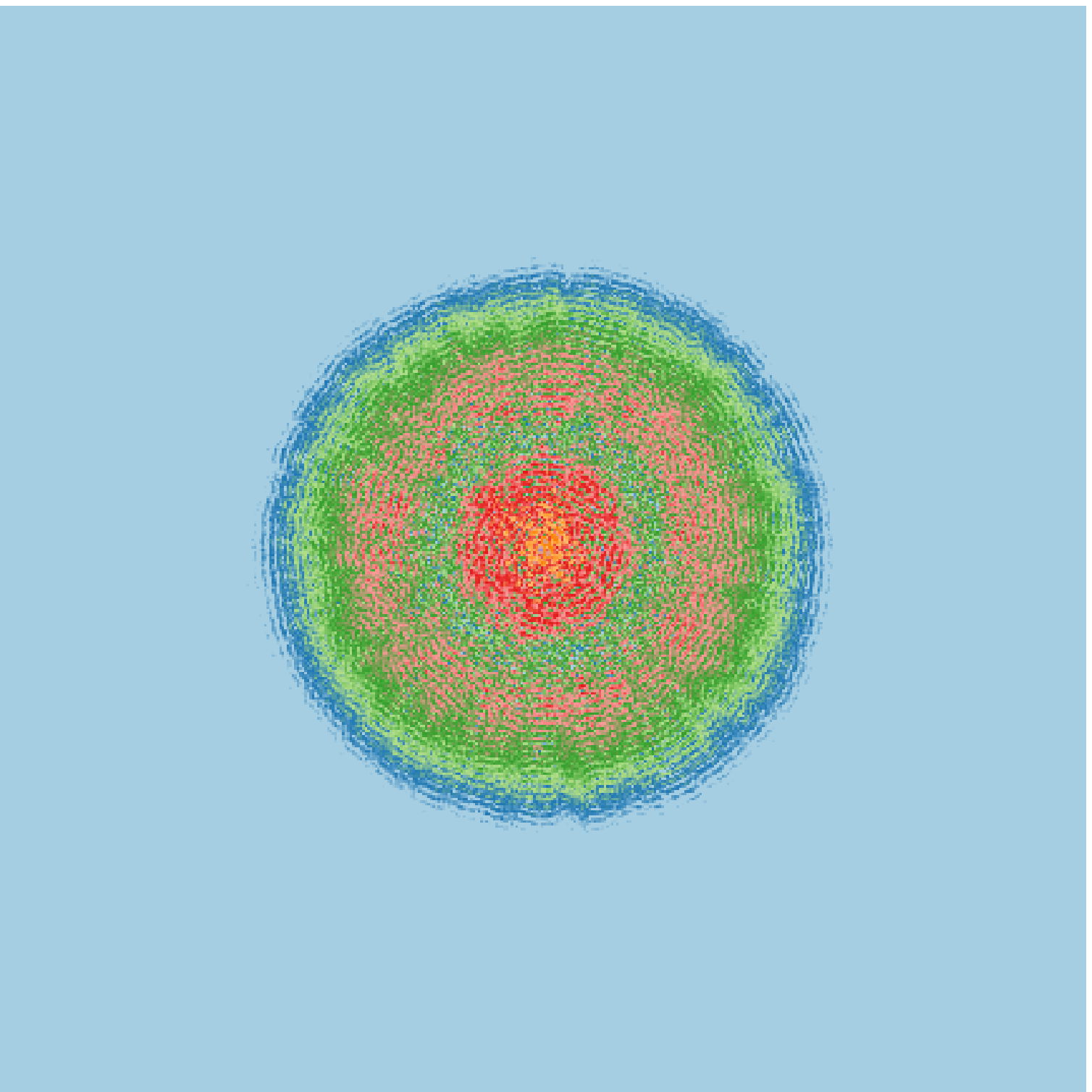}%
\includegraphics[width=0.165\textwidth]{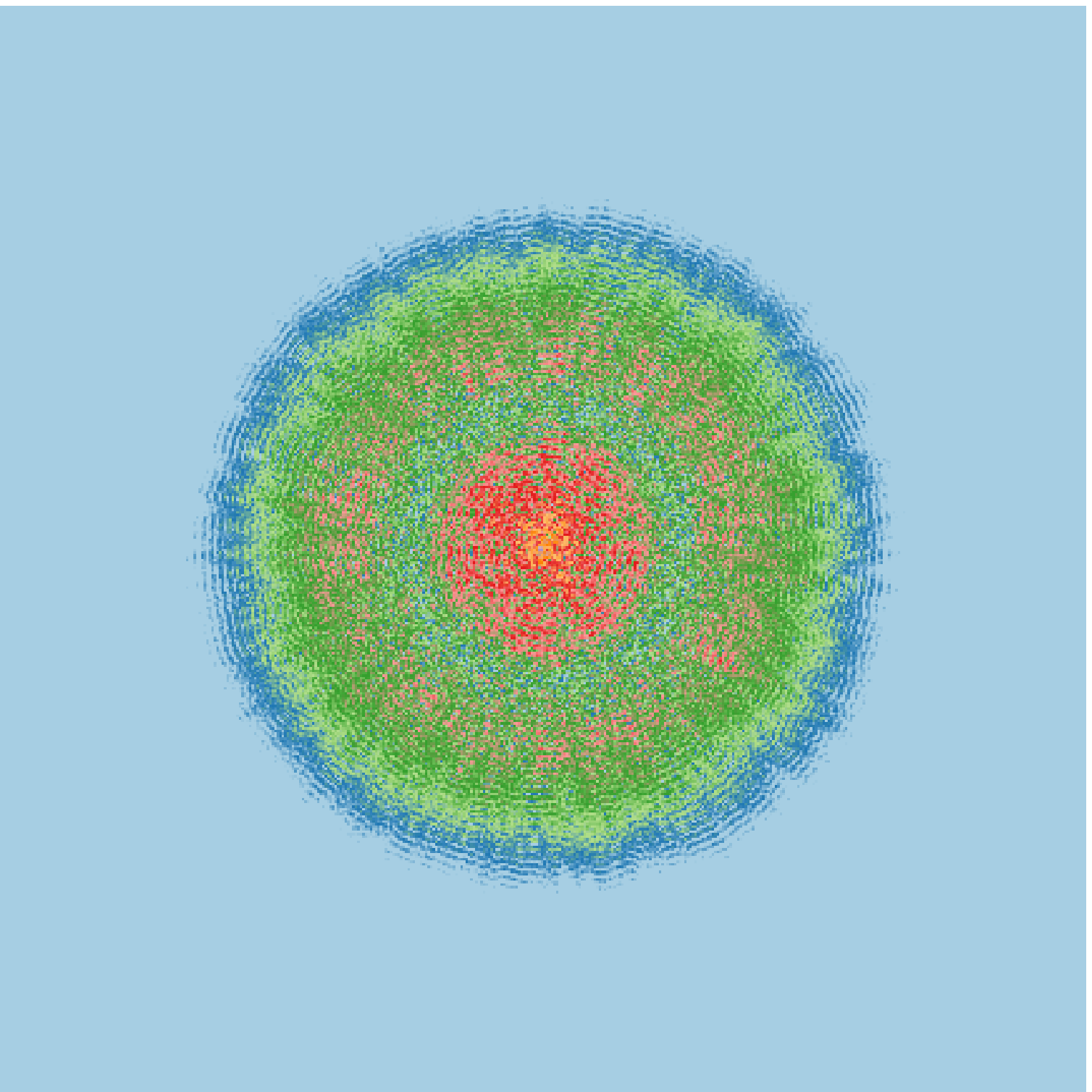}%
\includegraphics[width=0.165\textwidth]{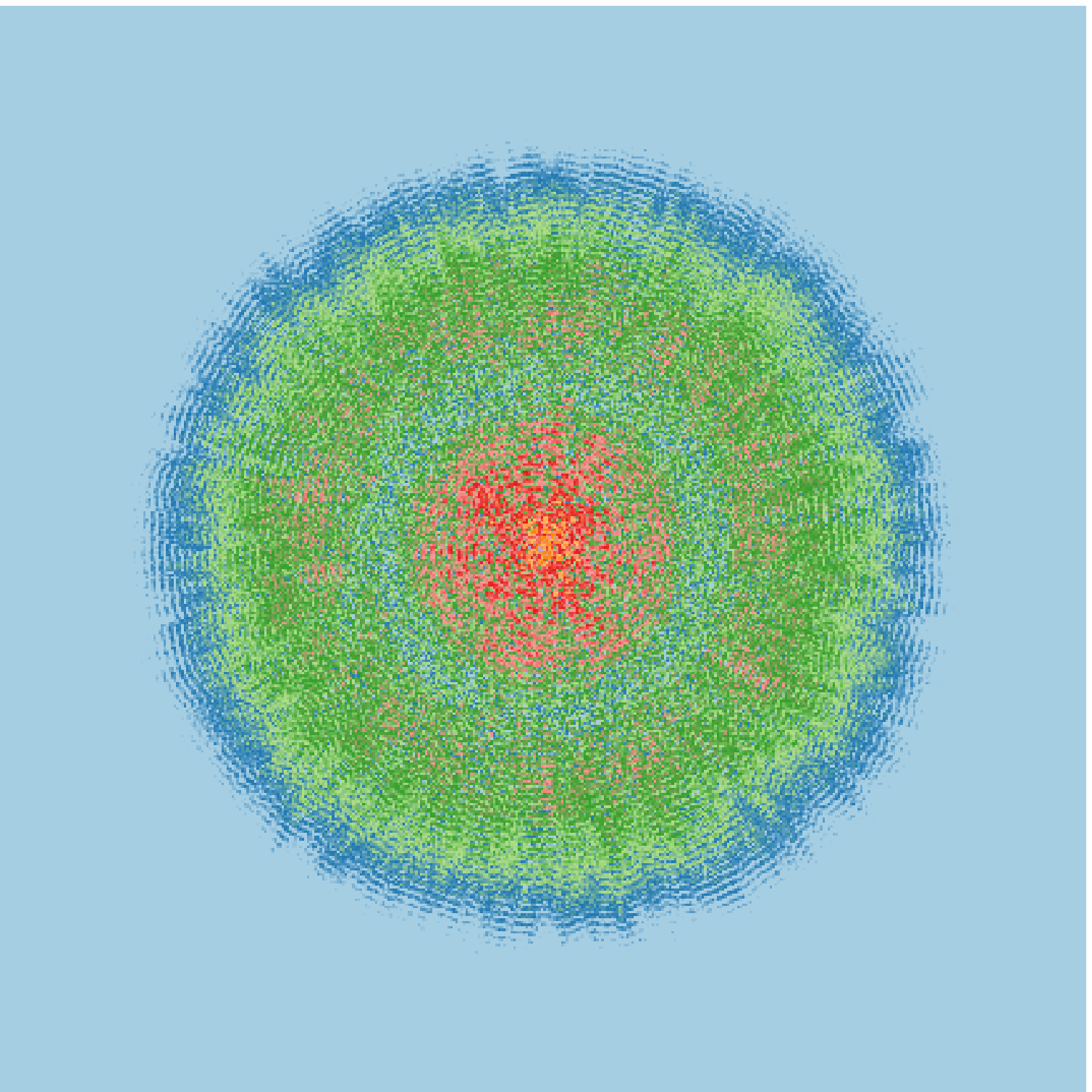}\\
\includegraphics[width=0.165\textwidth]{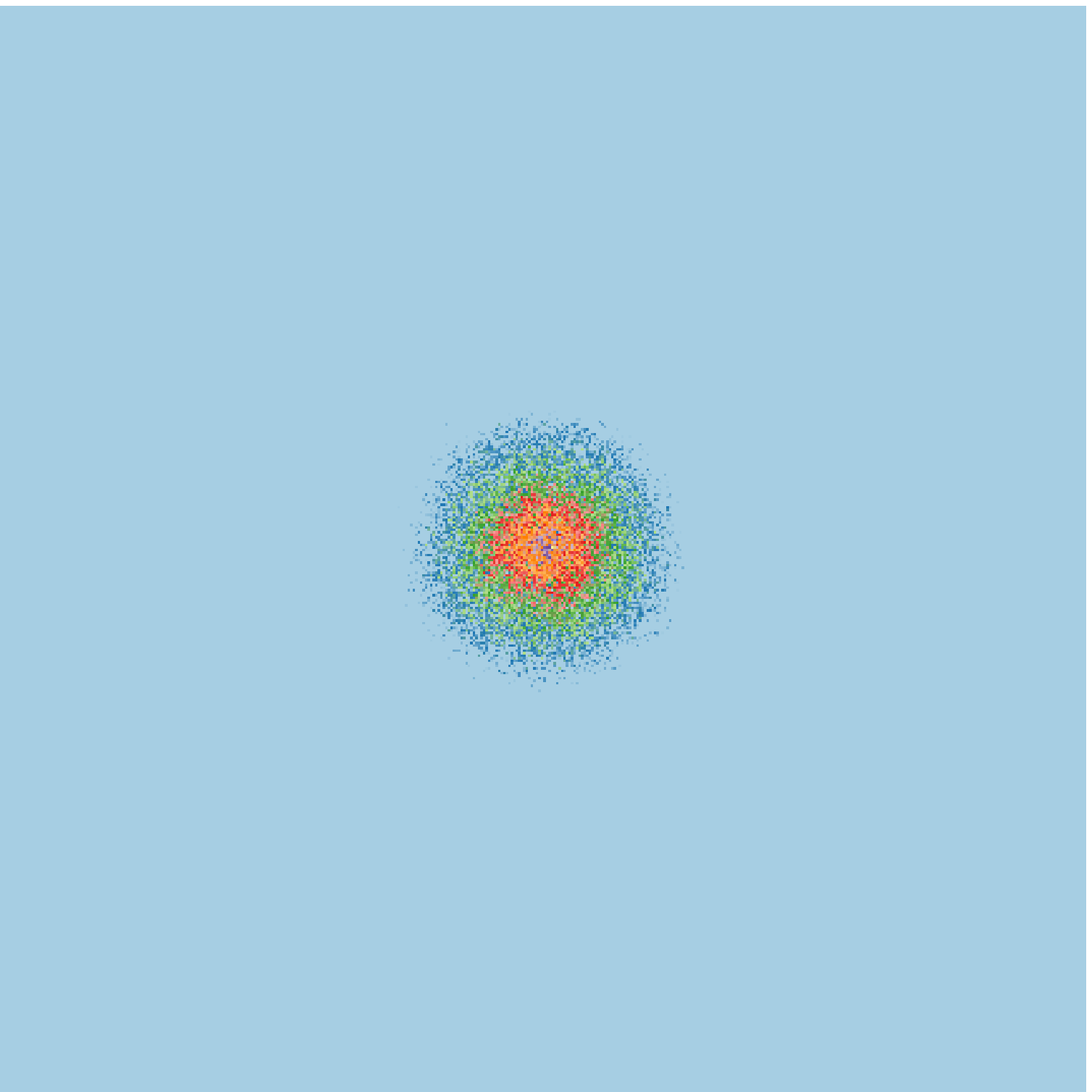}%
\includegraphics[width=0.165\textwidth]{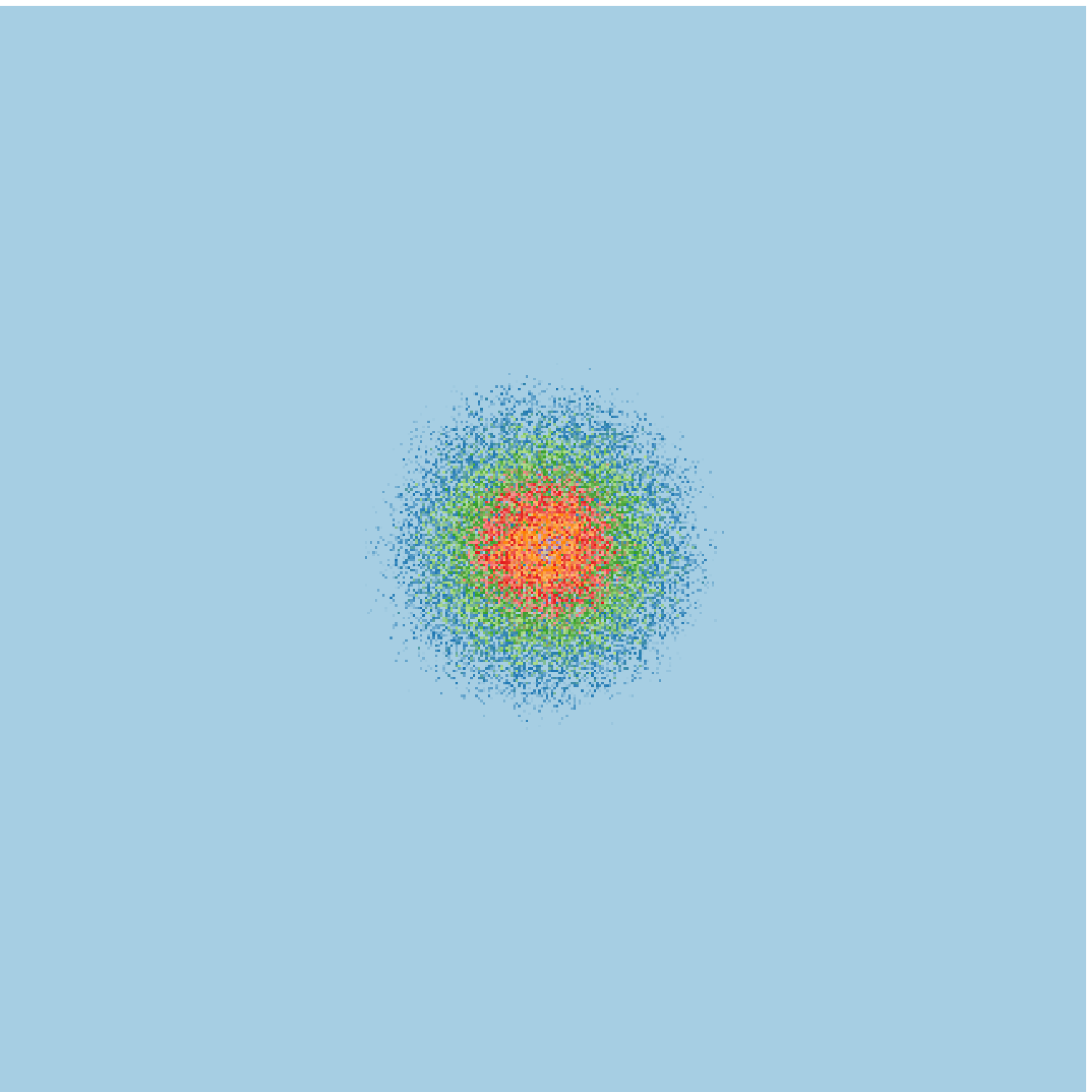}%
\includegraphics[width=0.165\textwidth]{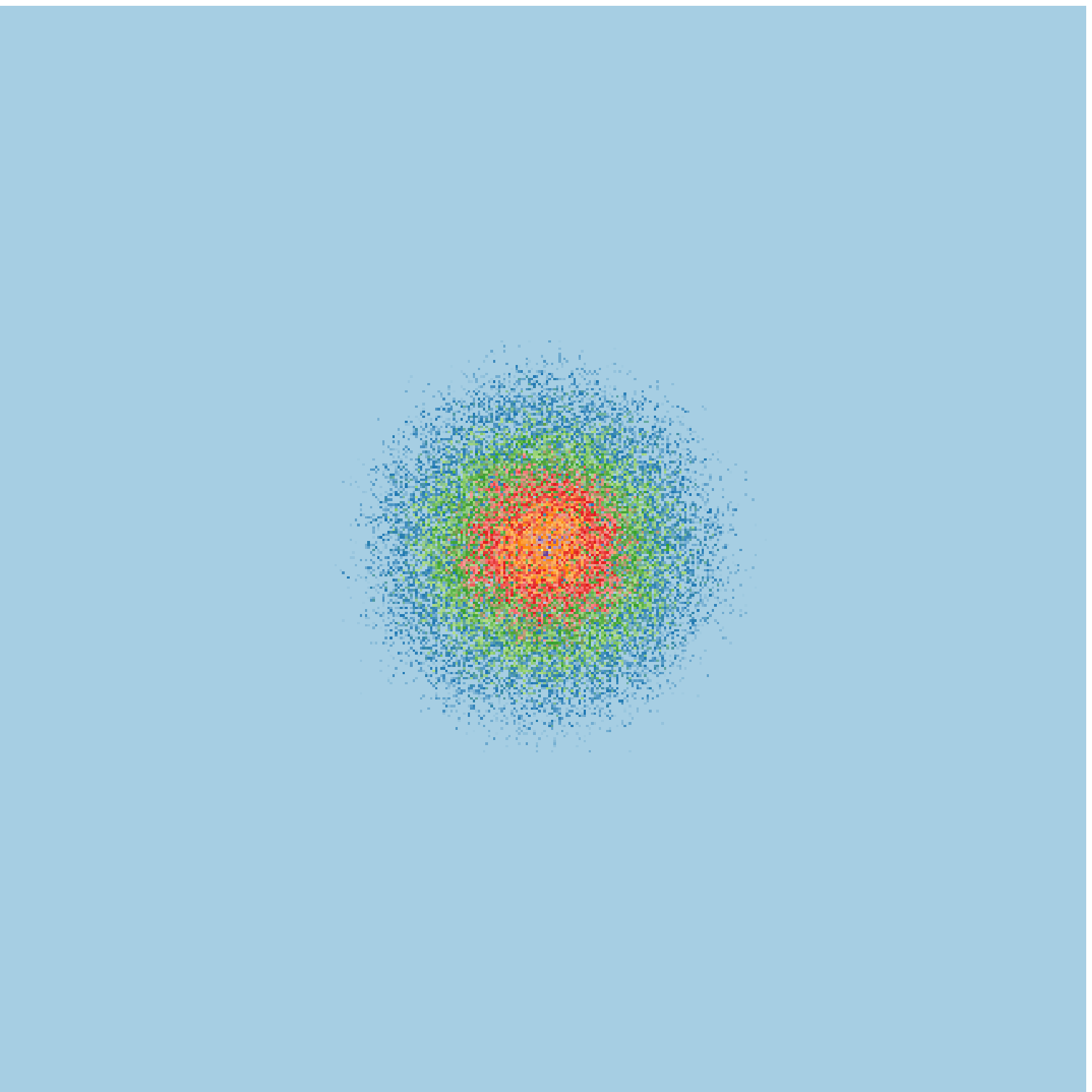}%
\includegraphics[width=0.165\textwidth]{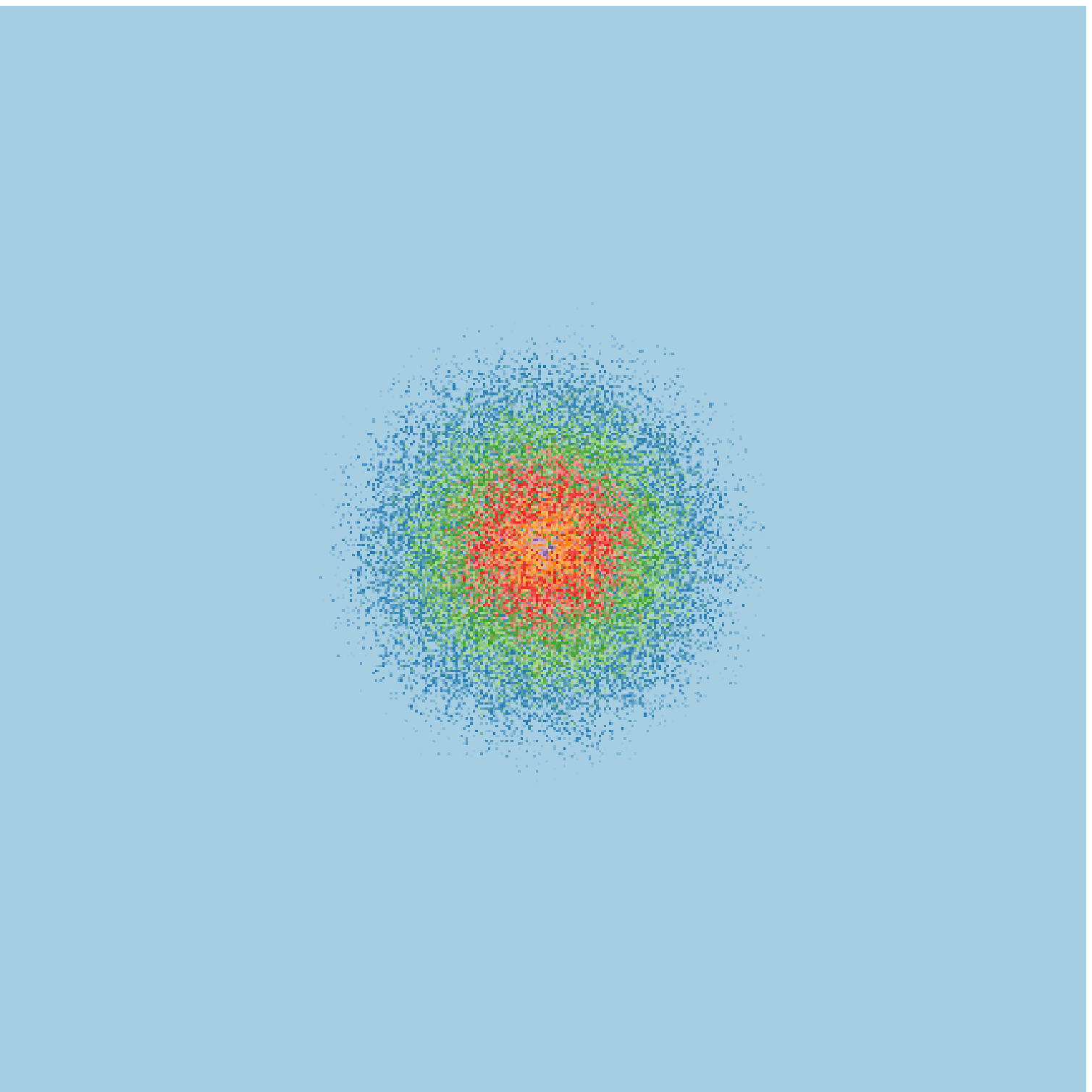}%
\includegraphics[width=0.165\textwidth]{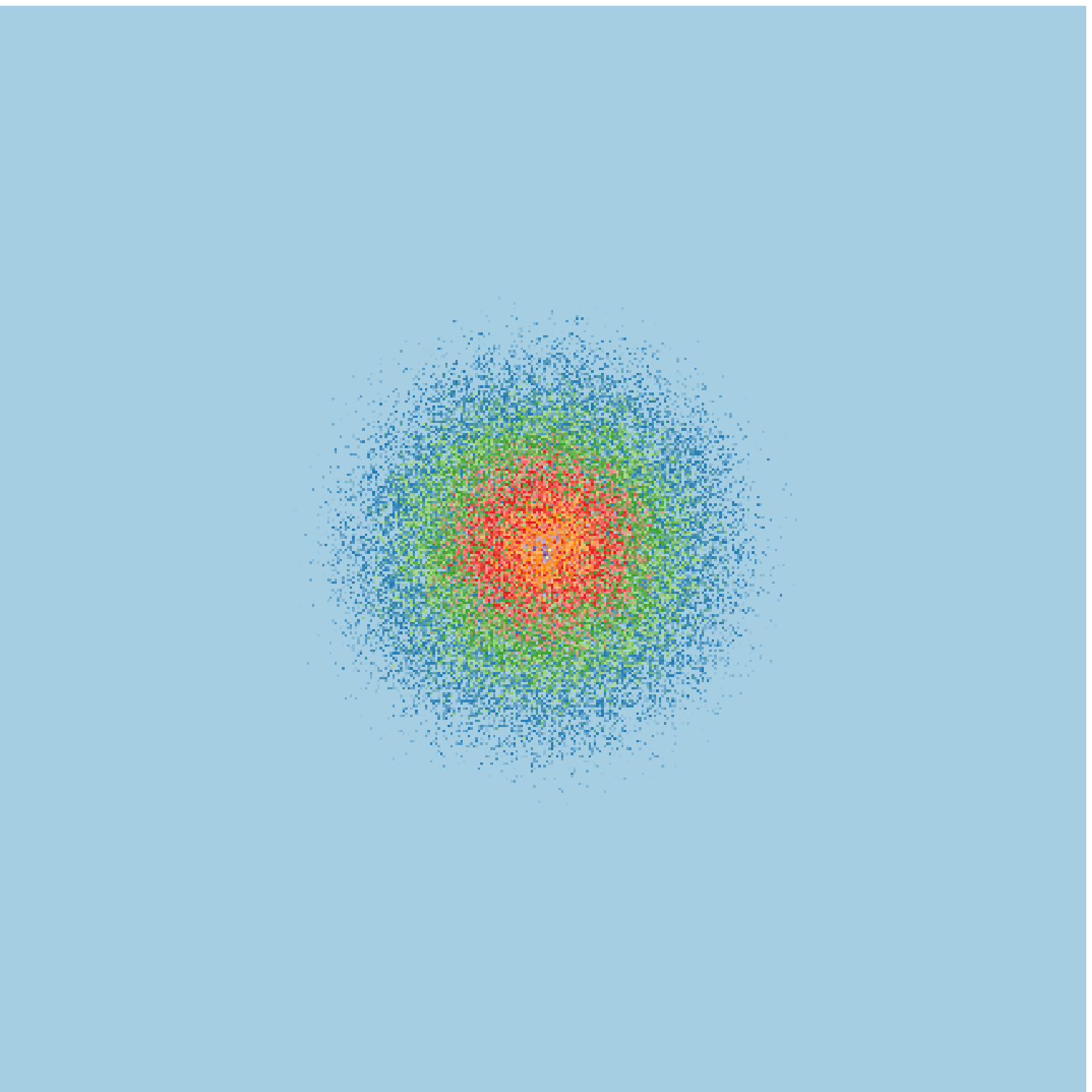}%
\includegraphics[width=0.165\textwidth]{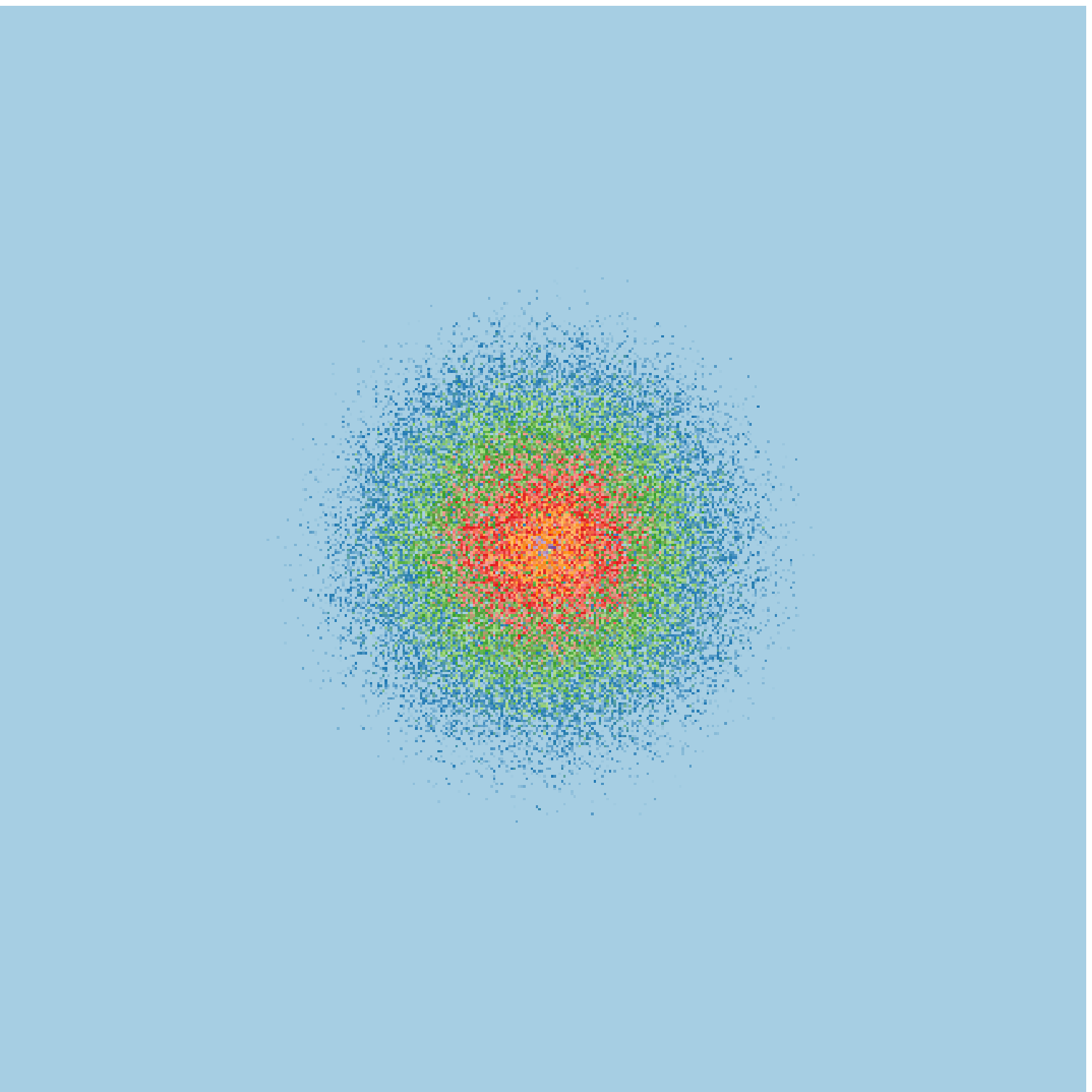}
\caption{(Color online) Time evolution of the wave function; we use a logarithmic scale \(\log|\psi_\uparrow(\bm x,t)|\), to visualize the small amplitude region. The lattice size is \(2048^2\). The exchange constant is \(J_s=0.1\) (top row), and \(J_s=1.0\) (bottom row); times are, from left to right, 160, 288, 416, 544, 672, 800. The concentration of magnetic impurities is \(c=0.1\), and the Rashba constant \(\lambda=0.2\). The color scale is fixed, and spans the interval \((10^{-4},5\,10^{-2})\).}
\label{f-psi}
\end{figure*}

The quantum state of the itinerant spins \(|\psi(t)\rangle\), evolves according to the Schrödinger equation 
\begin{equation}
\mathrm{i}\frac{\partial}{\partial t}|\psi(t)\rangle=H|\psi(t)\rangle\,,
\label{e-sch}
\end{equation}
which gives, in position representation, the spinor wave function \(\psi(\bm x,t)=\langle \bm x|\psi(t)\rangle\), where \(\bm x\) belongs to the square lattice, and which spin up and down components are \((\psi_{\uparrow},\psi_{\downarrow})\). We use the kernel polynomial method to integrate (\ref{e-sch}).\cite{Tal-Ezer-1984fk,Weisse-2008uq} The initial condition is a Gaussian wave packet of size \(d\) and unit norm,
\begin{equation}
\psi_{\uparrow}(\bm x,0) = \frac{1}{(2\pi d^2)^{1/2}}\exp\left(-\frac{|\bm x|^2}{4d^2}\right),
\label{e-gauss}
\end{equation}
which spin is in the up direction \(\psi=(\psi_{\uparrow},0)\). Most numerical computations were performed using \(L=2048\) and \(d=8\). Figure~\ref{f-psi} illustrates the evolution of the wave packet, for two values of the disorder strength.

The width of the wave packet is defined by, 
\begin{equation}
w^2_{\uparrow\downarrow}(t)=\int_{L^2}\!d\bm x \,|\bm x|^2 |\psi_{\uparrow\downarrow}(\bm x,t)|^2\,.
\label{e-w}
\end{equation} 
The asymptotic time dependence of the width, independent of the spin, satisfies a power law,
\begin{equation}
w(t) \sim t^\beta
\label{e-beta}
\end{equation}
characterized by the exponent \(\beta\); typical values are \(\beta=1\), for a free particle (the Rashba term do not change this linear dependence on time), and \(\beta=1/2\), in the diffusion regime (see Fig.~\ref{f-w}).

%
%
The geometry of the time dependent quantum state can be described by a measure based on the electron wave function \(\psi_i(t)\), which gives the probability to find the particle at site \(i\), \(|\psi_i(t)]^2\). Dividing the lattice into \(N_l\) cells of size \(l\), the dynamical measure of the \(k\) box is then given by,
\begin{equation}
m_k(l,t)=\sum_{i \in B_k(l)} \left|\psi_i(t)\right|^2,\quad k=1,\dots,N_l\,,
\label{e-mk}
\end{equation}
where we sum over the \(l^2\) sites of the \(B_k\) box. Using this measure we define the dynamical inverse participation ratios,
\begin{equation}
P_q(l,t)=\sum_{k=1}^{N_l}m_k(l,t)^q\,,
\label{e-pq}
\end{equation}
and the spectrum of fractal dimensions,
\begin{equation}
D_q(t)=\lim_{l\rightarrow0} \frac{1}{q-1} \frac{\log P_q(l,t)}{\log l}\,,
\label{e-dq}
\end{equation}
which determine, in the long-time limit, the asymptotic behavior,\cite{Huckestein-1994kx,Janssen-1994fk,Arias-1998kx}
\[
P_q(l) \sim l^{(q-1)D_q}\,.
\]
In the following we show results with \(q=2\), of the dynamical inverse participation ratio and correlation dimension, \(P_2(t)\) and \(D_2(t)\) (see Figs.~\ref{f-d2} and \ref{f-ipr}).

%
%
The local, spin dependent, density of states is defined by,
\begin{subequations}
\begin{eqnarray}
\rho_\uparrow(\bm x,\epsilon)&=&\sum_\alpha \cos^2(\theta/2)
|\psi_{\uparrow}(\bm x,\alpha)|^2 \delta(\epsilon-\epsilon_\alpha)\\
\rho_\downarrow(\bm x,\epsilon)&=&\sum_\alpha \sin^2(\theta/2)
|\psi_{\downarrow}(\bm x,\alpha)|^2 \delta(\epsilon-\epsilon_\alpha)
\end{eqnarray}
\label{e-rhoi}%
\end{subequations}
for the spin-up and spin-down components, respectively, where \(\theta\) is the angle of the impurity magnetic moment, chosen as the quantization axis, with respect to the \(z\) direction, \(\epsilon\) is the energy and \(\epsilon_\alpha\) is the eigenenergy of the eigenstate \(|\alpha\rangle\) (we use the notation \(\psi_{s}(i,\alpha)=\langle i|s,\alpha\rangle\), with  \(s=\uparrow, \downarrow\) is the spin index). The sum over the space gives the total density of states \(\rho(\epsilon)=\sum_i [\rho_\uparrow(i,\epsilon) + \rho_\downarrow(i,\epsilon)]\). As for the time dependent Schrödinger equation, we use the polynomial kernel method to compute (\ref{e-rhoi}).\cite{Weisse-2008uq,van-den-Berg-2011fk} This quantity is experimentally accessible, and shows a multifractal structure in the vicinity of the metal-insulator transition, on the surface states of a dilute magnetic semiconductor.\cite{Richardella-2010kl} The explicit dependence on the spin is useful for investigating the correlation of possibly localized electron spin states, with the spin of the impurities (see Fig.~\ref{f-ldos}).

%
\section{Wave packet dynamics and localization}
\label{S-wave}
%
%
In the usual Anderson model one considers a free particle that jumps between  sites having random energies, independently of its spin; in the present model, a particle jumps changing its spin, due to the spin-orbit coupling, and is scattered off by spatially distributed impurities. The study of the dynamics of an initially localized wave packet, is interesting for getting a qualitative image of the propagation of the quantum states in the field of randomly distributed scatters. We show in Fig.~\ref{f-psi}, the time evolution of (\ref{e-gauss}) for two values of the disorder, \(J_s=0.1\) in the top panel, and \(J_s=1\) in the bottom panel. Most numerical results are given for a concentration \(c=0.1\) of impurities, and a Rashba constant \(\lambda=0.2\). Not only the speed of propagation differs in both cases, but also the geometry of the wave function is qualitatively different. In the weak disorder case, one observes an interference pattern, and at long times, a star-like structure; for stronger disorder, the quantum probability density develops a granular and intermittent structure, with a slow decay of the amplitude at the origin.

\begin{figure}
\centering
\includegraphics[width=0.48\textwidth]{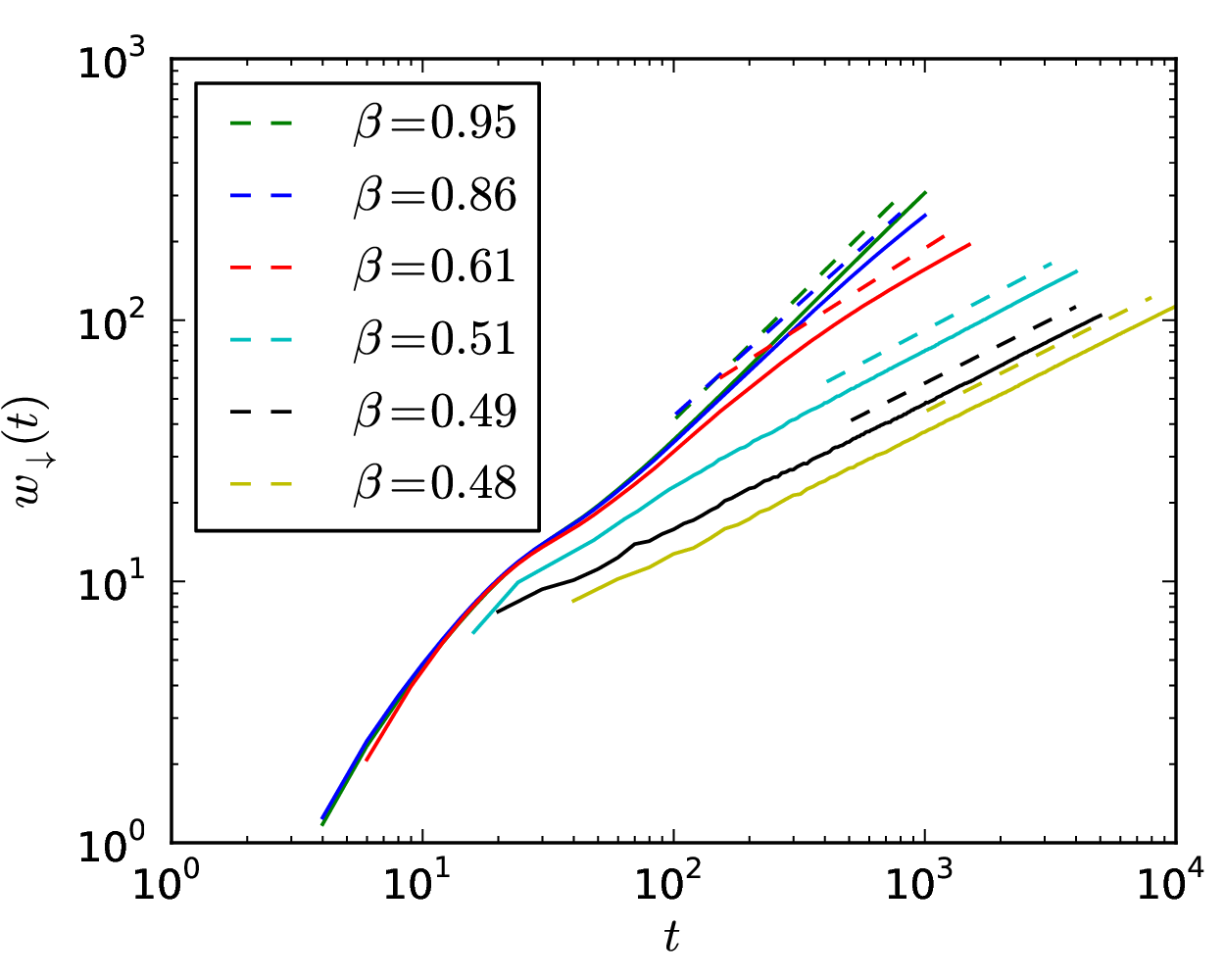}
\caption{(Color online) Width of the wave function showing a time power law with exponents dependoing on the disorder strength, from quasi free dispersion (\(\beta=0.95\), for \(J_s=0.1\)) to diffusion (\(\beta=0.51\) for \(J_s=1.0\)). Exchange constant \(J_s = 0.1\), 0.2, 0.4, 1.0, and 2.0 (from top to bottom).}
\label{f-w}
\end{figure}

%
%
These qualitative observations are confirmed more quantitatively by the measure of the wave packet's mean square displacement \(w(t)\), that characterizes its diffusion, and the correlation dimension, which is related to the decay exponent of the return probability. As shown in Fig.~\ref{f-w}, the asymptotic behavior of the width is well described by a power law (\ref{e-beta}), with an exponent \(\beta\) that decreases with increasing disorder. We find that for \(J_s \ll 1\) the motion is quasi ballistic, while for \(J_s \gtrsim 1\) a slightly subdiffusive regime sets in, as a possible manifestation of localization. The strong dependence of \(\beta\) on the disorder strength is reminiscent of the diffusion in a quasi periodic potential.\cite{Ketzmerick-1992fk,Cerovski-2005uq,Kolovsky-2011fk}

\begin{figure}
\centering
\includegraphics[width=0.48\textwidth]{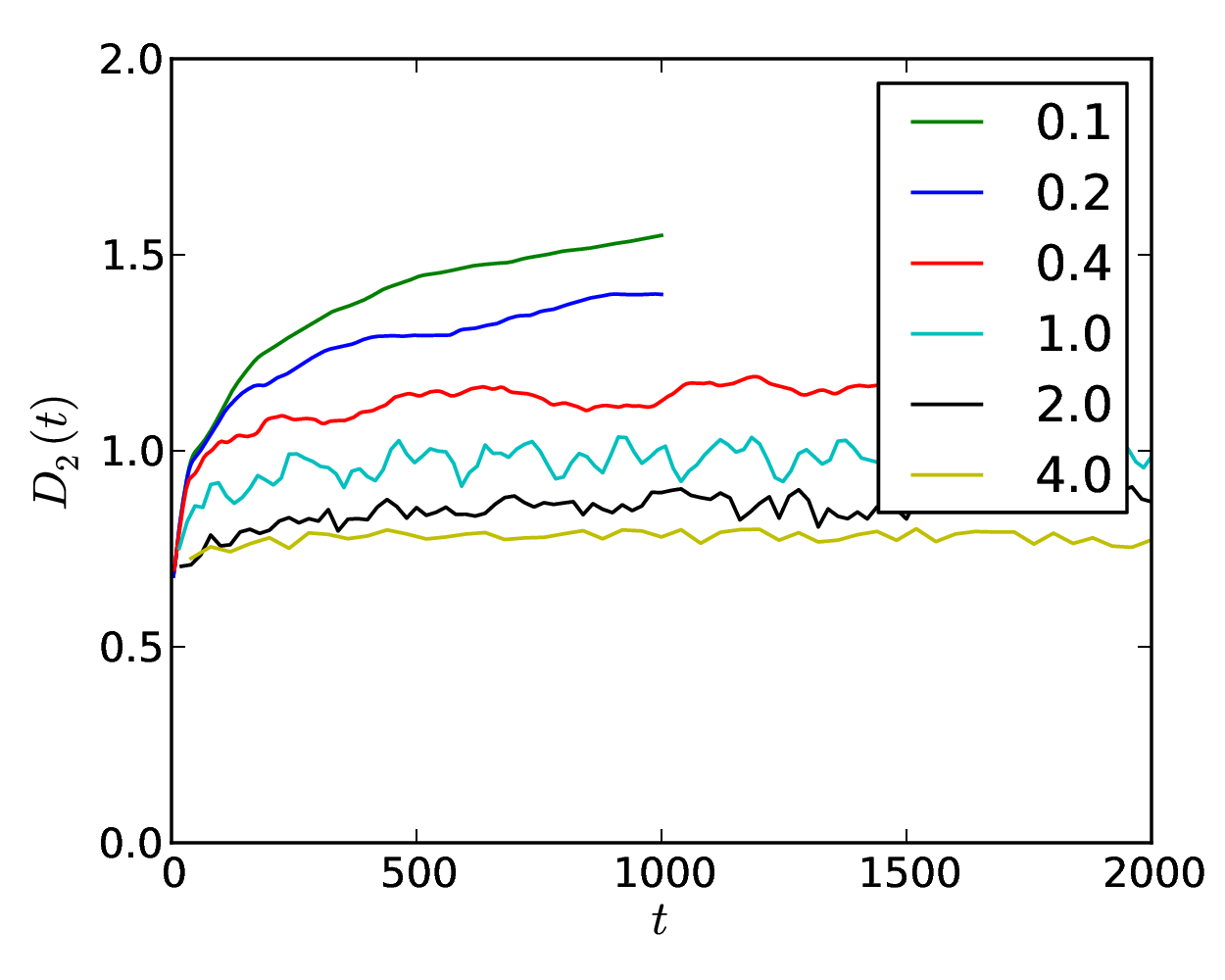}
\caption{(Color online) Measure of \(D_2\) as a function of time for different disorder strengths (\(J_s\) in the legend).}
\label{f-d2}
\end{figure}

%
%
The measure of the correlation dimension \(D_2\), which accounts for the number of sites visited at time \(t\),  \(w(t)^{D_2}\), also depends on the disorder strength. In Fig.~\ref{f-d2} we plot \(D_2(t)\) for \(J_s = 0.1\), 0.2, 0.4, 1.0, 2.0, and 4.0; the corresponding asymptotic values of the fractal dimension are \(D_2 = 1.45\), 1.32, 1.14, 0.99, 0.88, and 0.79. It is worth noting that for \(J_s=1\) we find \(\beta\approx0.5\) and  \(D_2\approx1\), these are just the exponents of a two-dimensional random walk, giving a strong indication of a critical state separating extended and localized regimes. For stronger disorder the correlation dimension passes below 1, \(D_2<1\), corresponding to a point-like support of the wave packet, and a persistence of the probability to return to the origin, in accordance with the phenomenology observed in Fig.~\ref{f-psi}.

\begin{figure}
\centering
\includegraphics[width=0.48\textwidth]{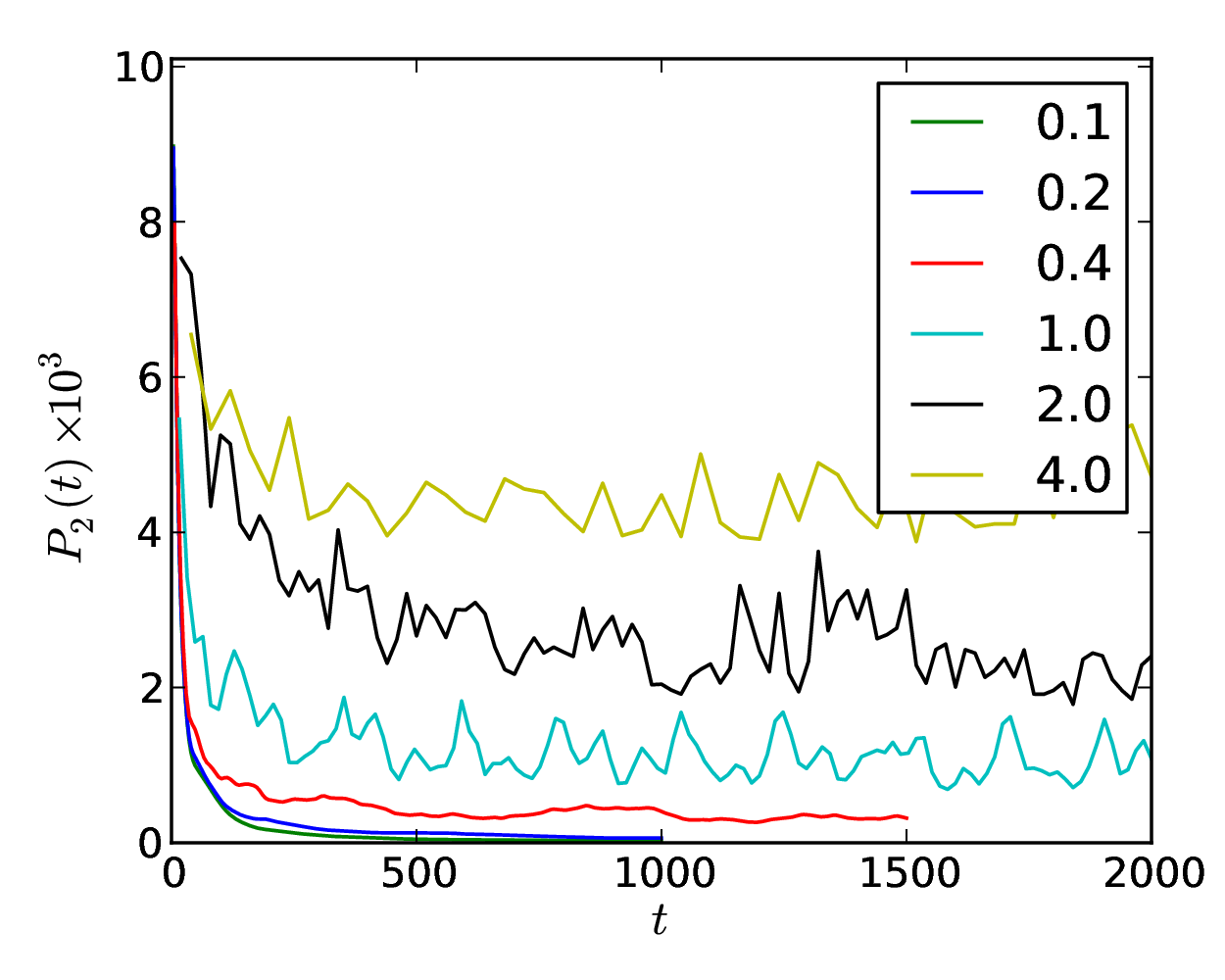}
\caption{(Color online) Inverse participation ratio \(P_2\) as a function of time for different disorder strengths (\(J_s\) in the legend). When \(J_s\gtrsim 1\), \(P_2(t)\) remains finite for long times, indicating the existence of a localization state.}
\label{f-ipr}
\end{figure}

%
%
A criterium to distinguish between extended and localized states is given by the asymptotic behavior of the dynamical inverse participation ratio \(P_2(t)\), that must vanish for extended states but remains finite for localized ones. We plot \(P_2(t)\) in Fig.~\ref{f-ipr}, using the same parameters as before, and remark that a visible change operates around \(J_s=1\); the asymptote is not only finite for \(J_s\gtrsim1\), but the temporal fluctuations are notably enhanced around the transition region. The picture shows a portion of the temporal evolution; for \(J_s=2\) and 4, the simulations run up to \(5\times10^3\) and  \(10^4\), respectively, without change in the mean value of \(P_2\). Therefore, from the dynamical point of view, a localization transition appears when the disorder is increased, around the values \(c=0.1\) and \(J_s=1\).

\begin{figure}
\centering
\includegraphics[width=0.48\textwidth]{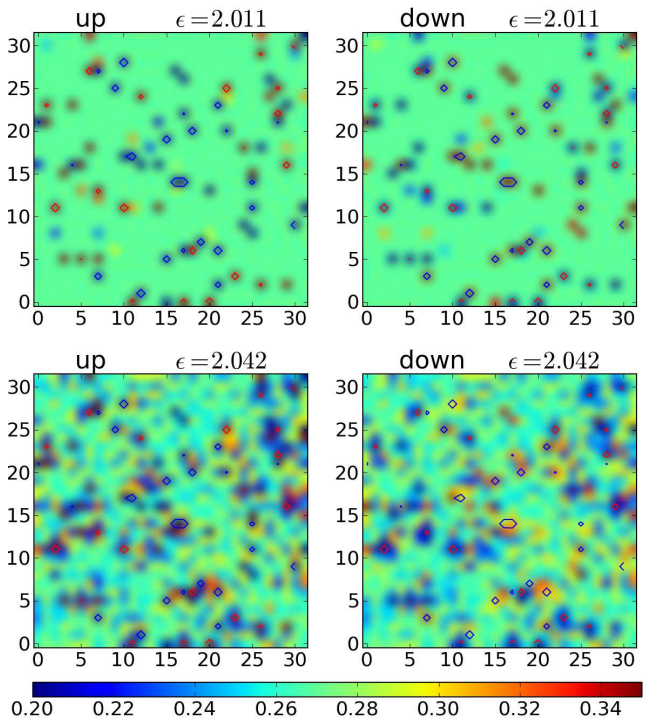}
\caption{(Color online) Spin up and down local density of states for \(J_s=0.1\) (top row) and \(J_s=1\) (bottom row). The contours (blue and red) are at the location of the impurities, where the \(z\) component of their magnetic moment is \(n_z=-0.5\) (blue) and \(-0.5\) (red). In the neighborhood of the impurities, at weak disorder strength, \(\rho_{\uparrow}(i)\) and \(\rho_{\downarrow}(i)\) compensate to give a value close to the one of the total density of states \(\rho\); at variance, for a stronger disorder, the spin of the quantum states is strongly correlated with the orientation of the impurity magnetic moment. The Fermi energy is around 2, and the bottom color bar gives the scale of \(\rho\); only a fraction \(32\times32\), of the total lattice is shown. }
\label{f-ldos}
\end{figure}

%
%
In order to complete the dynamical picture of the quantum system with its spectral properties, we investigate the local density of state (\ref{e-rhoi}). We find that for weak disorder the (spin independent) local density of states \(\rho(i)=\rho_\downarrow(i)+\rho_\uparrow(i)\), and the total one \(\rho\), coincide: there is no trace of strong inhomogeneities, for a given random position of the impurities. However, spin fluctuations are important even at weak disorder. Figure~\ref{f-ldos} presents the spin dependent local density of states, for the weak disorder case, upper panel \(J_s=0.1\), and for the transition region case,  bottom panel \(J_s=1\). When the disorder is weak, in the delocalized state, the spin fluctuations at the impurities sites compensate each other. Note in particular, the exchange of colors between the up-spin and down-spin images, around the impurities contours (top row). In contrast, for stronger disorder (bottom row), a depletion of the density of states at the spin-up locations is created, while at the spin-down locations, an enhancement is observed. In the incipient localized state, for parameters near to the transition, arises a strong correlation between the spin of the electrons and the direction of the magnetic moment of the impurities. This effect on the spin translates in a strong inhomogeneity of the spin independent \(\rho(i)\): total and local density of states no more coincide, and as in the usual Anderson transition, the arithmetic and geometric means of \(\rho(i)\) are no longer equal.

%
\section{Spin Hall conductivity}
\label{S-spin}
%
%
At weak disorder the spin Hall conductivity stays close to its clean value,\cite{van-den-Berg-2011fk} but already quantum interference effects play an important role (Fig.~\ref{f-psi}). This observation naturally leads to questions about the effect of interference on the spin Hall conductivity. Quantum corrections for a system with spin-orbit coupling and spin-independent scattering, were previously calculated for the charge conductivity in connexion with the anomalous Hall effect,\cite{Skvortsov-1998fk} as well as for the spin Hall conductivity.\cite{Chalaev-2005fk} Another question that arises from the investigation of the localization transition is that of the persistence of the spin Hall effect in the localized regime. The weak localization corrections, due to quantum coherent backscattering, are therefore of interest because they may give information about the quantum interference mechanisms that should also be relevant for spin transport in the strong disorder regime.

\subsection{Quantum corrections}

In order to understand quantum effects of the disordered magnetic impurities on the spin Hall effect we will calculate the first quantum corrections to the Kubo formula for the spin Hall conductivity using the parabolic continuous band approximation of the tight-binding Hamiltonian (\ref{e-H}). The continuous Hamiltonian is written
\begin{equation}
\label{Hcont}
H = \frac{{\bm p}^2}{2m} - \frac{\lambda}{\hbar} {\bm \sigma}\cdot(\hat{z}\times {\bm p}) +
    V(\bm x)\,,
\end{equation}
where \(\bm p\) and \(\bm x\) are the momentum and position operators of the electron in the plane \((x,y)\). The interaction potential is characterized as in the tight-binding model, by the exchange energy \(J_s\) and a characteristic microscopic length scale \(a\) (the lattice step in the tight-binding model). It can be written,
\begin{equation}
V(\bm{x})= J_s a^2\sum_{i \in \mathcal{I}} \begin{pmatrix}
       \cos \theta_i & \E^{- \I \phi_i}\sin \theta_i  \\
       \E^{\I \phi_i }\sin \theta_i & -\cos \theta_i 
\end{pmatrix} \delta(\bm{x} - \bm{x}_i)\,,
\end{equation}
where and \(\theta_i\) and \(\phi_i\) are the angles of the magnetic moment of an impurity located at \(\bm{x}_i\). The concentration of impurities will be denoted by \(c\). We work in nondimensional units, such that \(\hbar = m = a = e = 1\). The spin splitting at the Fermi energy is given by \(\Delta=2\lambda k_F\), where \(k_F=\sqrt{2\epsilon_F}\) is the Fermi wavenumber. The clean Hamiltonian is diagonal in the chiral basis, with eigenenergies 
\begin{equation}
\epsilon_{\pm} (k) = \frac{k^2}{2} \pm \lambda k\,,
\end{equation}
where \(k\) is the modulus of the wavevector \(\bm{k}=(k_x,\,k_y) = k\,(\sin \varphi,\,\cos \varphi)\).

\begin{figure}
\centering
\includegraphics[width=0.46\linewidth]{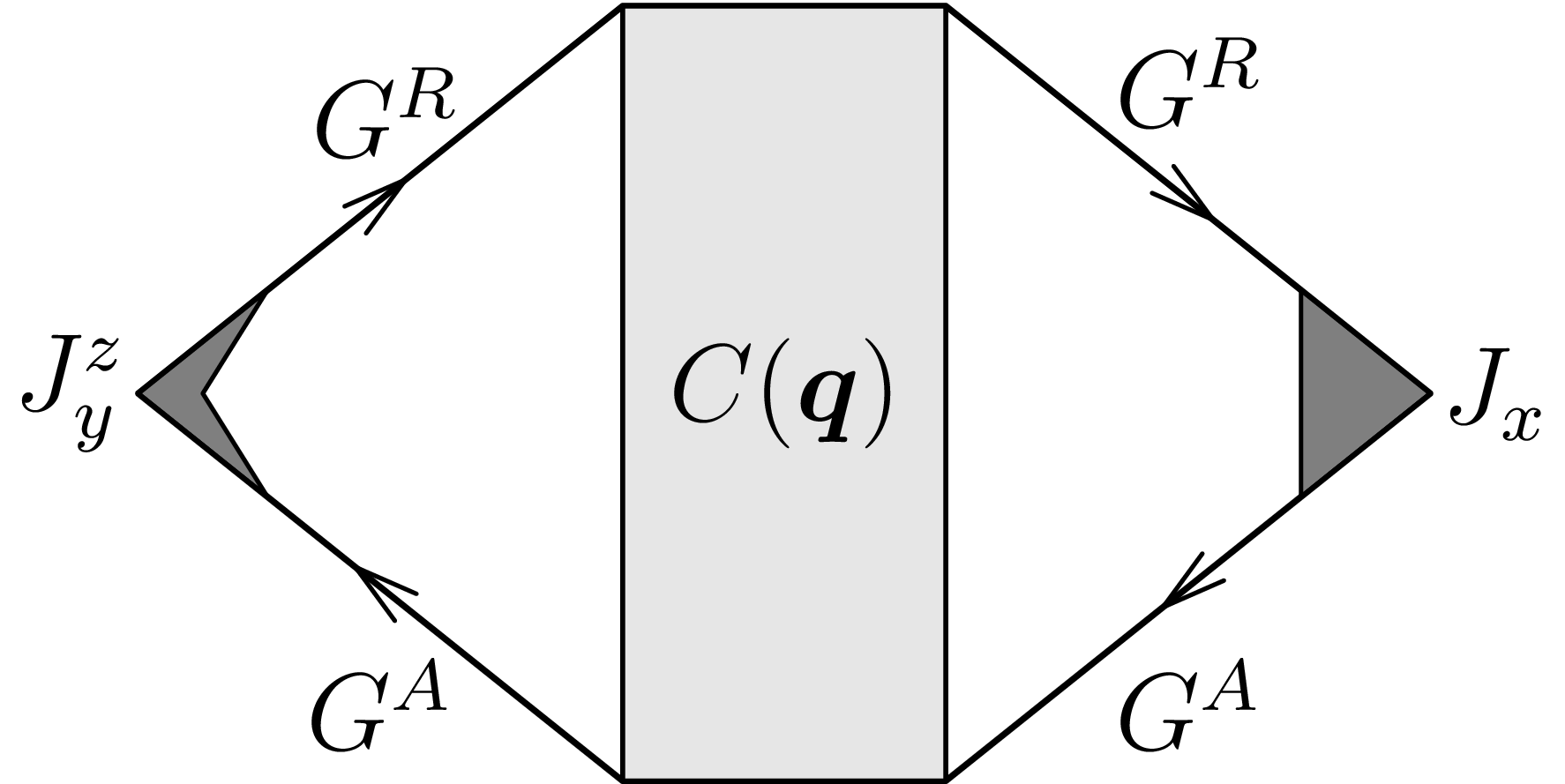}(1)\\[1em]
\includegraphics[width=0.46\linewidth]{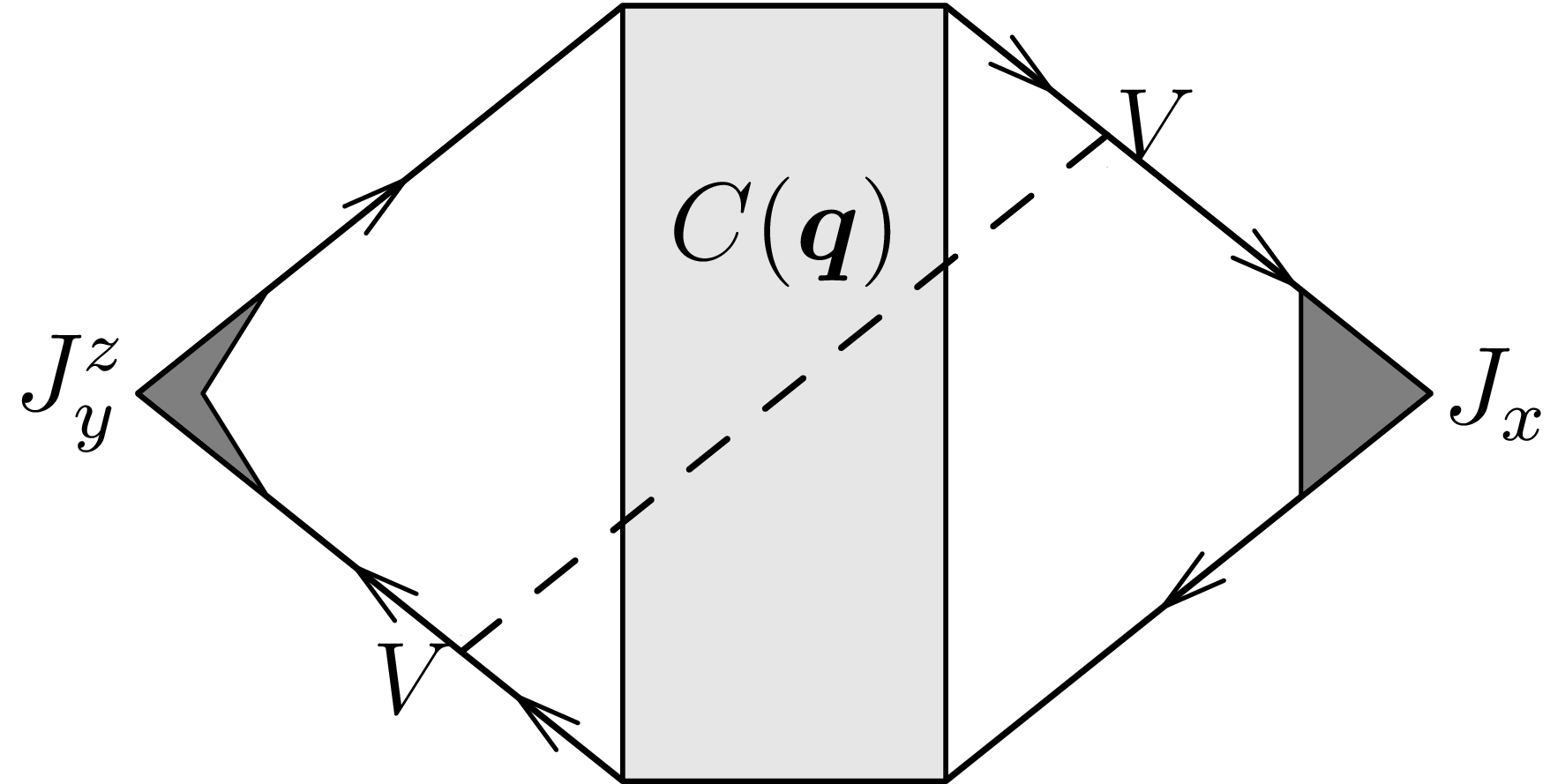}(2)%
\includegraphics[width=0.46\linewidth]{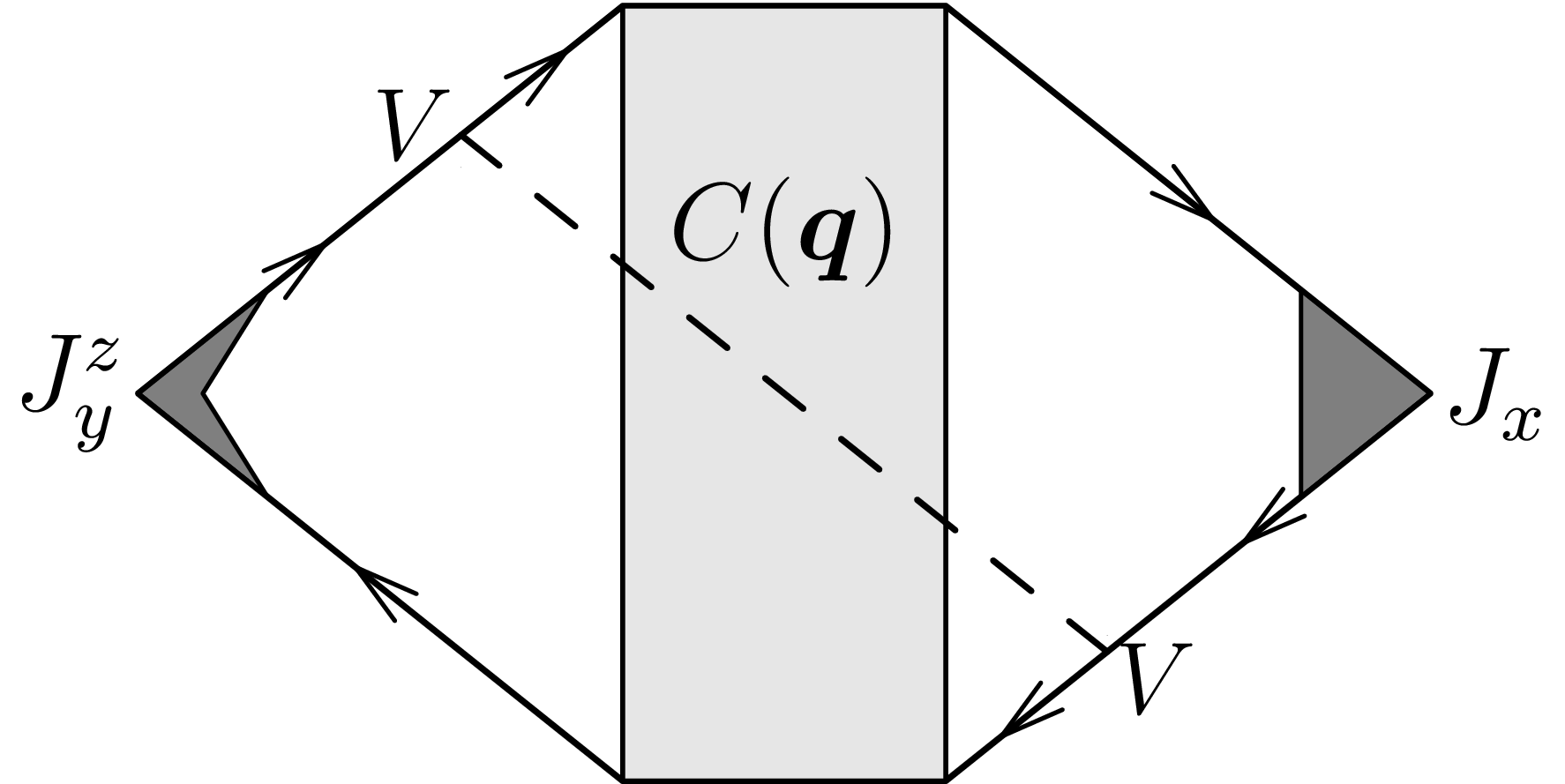}(3)%
\caption{\label{f-hikami}The three Hikami boxes for the weal localization corrections to the spin conductivity. The spin and charge current vertex are renomalized.}
\end{figure}

The linear (spin Hall) response of the system, to an applied electric field \(E_x\), is given by the Kubo formula for the spin Hall conductivity. It is convenient to express the Kubo formula in terms of retarded \(G^R\), and advanced \(G^A\), Green functions, as\cite{Dimitrova-2005uq,Chalaev-2005fk}
\begin{align}
\label{eq.sigma(G)}
\sigma^z_{xy} (\omega) = & \frac{1}{\omega} 
    \left\langle\mathrm{Tr} \left[ 
    j_y^{z} (\bm{k}) G^R (\epsilon+\omega,\bm{k}) 
    j_x (\bm{k}) G^< (\epsilon, \bm{k}) \right. \right.  
\nonumber \\
    & + \,\left. \left. 
    j_y^{z} (\bm{k}) G^< (\epsilon+\omega,\bm{k}) 
    j_x (\bm{k}) G^A(\epsilon, \bm{k}) 
    \right] \right\rangle,
\end{align}
where 
\[
\mathrm{Tr}[\dots]=\int \frac{d^2 \bm{k}}{(2 \pi)^2} 
    \int \frac{d\epsilon}{2 \pi} \mathrm{Tr}_s \,(\dots)
\]
\(\mathrm{Tr}_s\) is the trace over the spin index, \(\langle\dots\rangle\) denotes the average over the spatial disorder and magnetic moment orientations, and \( G^< (\epsilon) = f(\epsilon) (G^R(\epsilon)-G^A(\epsilon)) \), where \(f\) is the Fermi function. 

The charge current operator is \(j_x=-v_x=- \I\,[H,x]\) and the spin current operator is \(j^z_y=(1/4)\{\sigma_z,v_y\}\). The disorder averaged retarded and advanced Green functions are written~\cite{Skvortsov-1998fk}
\begin{equation}
\label{eq.Green1}
 G^{R,A}(\epsilon,\bm{k}) = \frac{\epsilon - \frac{\bm{k}^2}{2} + \lambda {\bm \sigma}\cdot(\hat{z}\times {\bm k} )}{(\epsilon - \epsilon_+(k) \pm \frac{\I}{2\tau})(\epsilon - \epsilon_- (k) \pm \frac{\I}{2\tau})}\,,
\end{equation}
where the spin-flip time \(\tau\) is defined by the self-energy as 
\begin{equation}
\frac{1}{\tau} =2 \pi N_F c  J_s^2 \approx c J_s^2\,.
\end{equation}

\begin{figure}
\centering%
\includegraphics[width=0.78\linewidth]{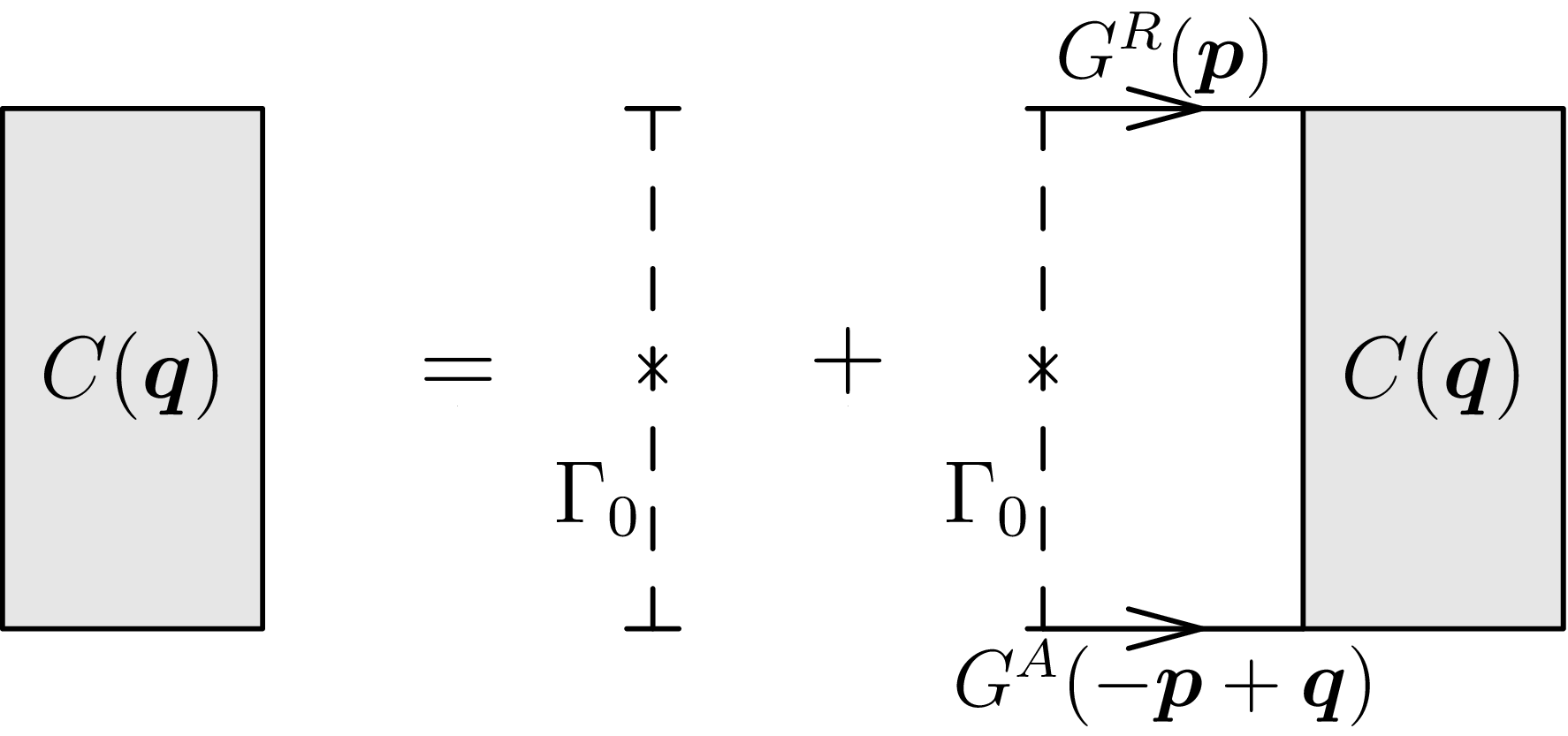}%
\caption{\label{f-cooperon}Diagrammatic representation of the Cooperon equation. }
\end{figure}

We assume that the spin splitting is small and disorder weak to that the Fermi energy is the largest energy scale of the system. This leads to the approximation \(\epsilon_F \gg \Delta, \tau^{-1}\) and \(N_F \approx 1/(2\pi)\). The nondimensional parameter \(x=\Delta \tau\) contains information about the spin-splitting and the disorder strength, and can take arbitrary positive values. Using this parameter, the Green functions can be written in a nondimensional expression as
\begin{equation}
\label{eq.Green2}
 G^{R,A}(\xi,\varphi) = \frac{-\xi(k) \pm  \I + x ( \sigma_x\,\sin \varphi - \sigma_y\,\cos \varphi )}{(-\xi(k) - x \pm + \I)(-\xi(k) + x \pm \I)} \,,
\end{equation}
where \(\xi=2\tau (k^2/2-\epsilon_F)\). The fundamental interaction vertex \(\Gamma_0\) is written in terms of the impurity concentration and the interaction term as 
\begin{align}
\label{eq.Gamma0}
\Gamma_0\ &= c \int_{\Omega} \, V (\theta, \phi) \otimes V (\theta, \phi)\,, \nonumber \\
& =\frac{1}{(2 \pi N_F) 3 \tau}
\begin{pmatrix}
1&0&0&0\\
0&-1&2&0\\
0&2&-1&0\\
0&0&0&1\\
\end{pmatrix}\,,
\end{align}
where the four center elements are non-zero due to spin mixing.

For convenience we write all expressions in matrix notation, using the following definitions:
\[
A_{\alpha \beta} B_{\gamma \delta} = \mathcal{C}^{\alpha \beta}_{\gamma \delta} \rightarrow A \otimes B = \mathcal{C} 
\]
is the definition of the Kronecker tensor product and 
\[
\left( \mathcal{A} \mathcal{B}\right)^{\alpha \alpha '}_{\beta \beta '} = \sum_{\delta, \gamma} \mathcal{A}^{\alpha \gamma}_{\beta \delta} \mathcal{B}^{\gamma \alpha '}_{\delta \beta '} \rightarrow  \mathcal{A} \mathcal{B}
\]
is the definition of matrix multiplication. For the sake of clarity we will note \(4\times 4\) matrices using a calligraphic letter style. 

The spin Hall conductivity in the linear response framework, can be computed as a perturbation series in the disorder strength,
\begin{equation}
\sigma_{sH} = \sigma_{sH}^0 + \sigma_{sH}^L + \delta \sigma_{sH}\,,
\label{eq.sH}
\end{equation}
where the first order takes into account the zero-interaction-loop spin conductivity (equivalent to the Drude term for the electrical conductivity), and the other ones contain the classical ladder corrections, plus lowest order quantum corrections. The first two terms in (\ref{eq.sH} were computed in Ref.~\onlinecite{van-den-Berg-2011fk}; here we compute the third term  (maximally crossed diagrams).

The terms in the perturbation series with the quantum corrections \(\delta \sigma_{sH}\),  can be diagrammatically represented by the so-called Hikami boxes, as shown in Fig.~\ref{f-hikami}.\cite{Gorkov-1979ve,Hikami-1980fk} The total first order quantum corrections are the sum of the contributions of the three boxes \(\delta \sigma_{sH} = \delta \sigma^{(1)}_{sH} + \delta \sigma^{(2)}_{sH} +\delta \sigma^{(3)}_{sH} \).

Using the diagrams of Fig.~\ref{f-hikami}, we obtain the following expressions for the three Hikami boxes:
\begin{widetext}
\begin{align}
\delta \sigma^{(1)}_{sH} = - \int_{\bm{q}}  \int_{\bm{k}} &(G^A  (\bm{k}) J_y^z G^R(\bm{k}) ) \, \otimes (G^R (\bm{q}-\bm{k}) J_x G^A(\bm{q}-\bm{k}) )\, \mathcal{C} (\bm{q})\,, 
\label{eq.hikami1}\\
\delta \sigma^{(2)}_{sH} = - \int_{\bm{q}} \int_{\Omega} & \int_{\bm{k}} \int_{\bm{k'}}  (G^A(\bm{k})  J_y^z(\bm{k}) G^R(\bm{k}) V (\theta, \phi) G^R(\bm{k'})) \nonumber \\
&\qquad \otimes (G^R (\bm{q} -\bm{k'},) J_x(\bm{q}-\bm{k'}) G^A(\bm{q} -\bm{k'}) V (\theta, \phi) G^A(\bm{q} -\bm{k}) )\, \mathcal{C} (\bm{q}) \,, 
\label{eq.hikami2} \\
\delta \sigma^{(3)}_{sH} = - \int_{\bm{q}} \int_{\Omega} &\int_{\bm{k}} \int_{\bm{k'}} (G^A(\bm{k}) V (\theta, \phi) G^A(\bm{k'}) J_y^z(\bm{k'}) G^R(\bm{k'})) \nonumber \\ 
&\qquad \otimes (G^R (\bm{q} -\bm{k'}) V (\theta, \phi) G^R(\bm{q}-\bm{k}) J_x(\bm{q}-\bm{k}) G^A(\bm{q} -\bm{k}))\, \mathcal{C} (\bm{q})\,,
\label{eq.hikami3}
\end{align}
\end{widetext}
where we used the notations \( \int_{\bm{q}} = \int d\bm{q}/(2\pi)^2 \), \(\int_{\Omega} =\int d\theta \,d\phi \sin \theta/4\pi\) is the integration over all magnetic moments configurations. In (\ref{eq.hikami1})-(\ref{eq.hikami3}), \(J_x\) is the renormalized charge current vertex
\begin{equation}
J_x =\frac{-\Delta}{\sqrt{8 \epsilon_F}} \frac{(2+x^2)}{8+7x^2} \, \sigma_y \,,
\end{equation}
where we remark the appearance of an extra \(\Delta\) factor, and \(J_y^z\) is the renormalized spin current vertex of \(z\)-spins in the \(y\)-direction
\begin{equation}
J_y^z =  - \frac{kx}{2(8+7x^2)} \, \sigma_y \,.
\end{equation}
Taking into account the renormalization of the vertices turns out to be important, as they will rescale the quantum corrections to the spin conductivity with a factor proportional to \(\Delta/\epsilon_F\). This correction originates in the anomalous contribution to the velocity operator due to the spin-orbit coupling.

\(\mathcal{C}\) is the Cooperon, a \(4\times 4\) matrix coupling the incoming to the outgoing spin states. Physically it gives the amplitude of corrections due to quantum interference of counter propagating trajectories. The Cooperon satisfies the self-consistent Bethe-Salpeter equation, that is most easily expressed in terms of the diagram in Fig.~\ref{f-cooperon}.
In order to derive the spin conductivity we need the Cooperon at zero frequency (static) and at the Fermi energy; in this approximation it satisfies the following equation,
\begin{equation}
\mathcal{C}(\bm{q}) = \Gamma_0\ + \Gamma_0\ \int_{\bm{p}} G^R(\bm{p}) \otimes G^A(-\bm{p}+\bm{q}) \, \mathcal{C}(\bm{q})\,.
\end{equation}
The derivation of the explicit expression of the Cooperon is presented in the Appendix. The \(\bm{q}\)-dependence of its elements can be schematically expressed in the form [see Eq.~(\ref{eq.coopq})]
\begin{equation}
\mathcal{C} (\bm{q})  \sim \frac{a(x,\E^{\I \chi})}{2 \pi N_F \tau} \frac{(b(x) + l^2 q^2)}{(c(x) + l^2 q^2)(d(x) + l^2 q^2)}\,,
\end{equation}
where \(a,b,c\) and \(d\) are nondimensional rational expressions of \(x\) and \(\E^{\I\chi}\), and \(\chi\) is the angle between \(\bm{k}\) and \(\bm{q}\); \(l = \tau k_F\) is the mean free path. From the explicit expression (\ref{eq.coopq}) we obtain, the behavior of the Cooperon in the large \(x\) limit,
\begin{align}
&\lim_{x \rightarrow \infty} C(\bm{q}) =  \frac{1}{2 \pi N_F \tau}  \times \nonumber \\&
\begin{pmatrix}
 \frac{16 \left(l^2 q^2 +10\right)}{3 l^4 q^4+80 l^2 q^2+400} & 0 & 0 & \frac{-8 e^{-2 i \chi } l^2 q^2}{3 l^4 q^4 +80 l^2 q^2+400} \\
 0 & \frac{l^2 q^2+2}{6 \left(l^2 q^2-4\right)} & \frac{l^2 q^2-10}{6 \left(l^2 q^2-4\right)} & 0 \\
 0 & \frac{l^2 q^2-10}{6 \left(l^2 q^2-4\right)} & \frac{l^2 q^2+2}{6 \left(l^2 q^2-4\right)} & 0 \\
 \frac{-8 e^{2 i \chi } l^2 q^2}{3 l^4 q^4+80 l^2 q^2+400} & 0 & 0 & \frac{16 \left(l^2 q^2+10\right)}{3 l^4 q^4+80 l^2 q^2+400}
\end{pmatrix}
\end{align}
For small values of \(x\) we get,
\begin{align}
&\lim_{x \rightarrow 0} C(\bm{q}) =  \frac{1}{2 \pi N_F \tau}  \times \nonumber \\&
\begin{pmatrix}
\frac{2}{4+l^2q^2} & 0 & 0\\
0 & \frac{2l^2q^2}{-16+l^2q^2} & \frac{8}{16-l^2q^2} & 0\\
0 & \frac{8}{16-l^2q^2} & \frac{2l^2q^2}{-16+l^2q^2} & 0\\
0 & 0 & 0 & \frac{2}{4+l^2q^2}
\end{pmatrix}\,.
\end{align}
From the above two expressions we note that the \(q\rightarrow0\) limit of the Cooperon, changes according to the value of \(x\). This implies that we must retain the full dependence on \(x\) in order to obtain a reliable result for the spin conductivity. Indeed, in the calculation of the spin Hall conductivity we are interested in the \(\bm{q}\) close to zero limit, that allows us to simplify the \(\bm{q}\)-dependence of the Green functions and to factor out, in (\ref{eq.hikami1})-(\ref{eq.hikami3}), the \(q\) integration of the Cooperon. Because the interaction is isotropic on average (paramagnetic state), the two elements containing the angle \(\chi\) will vanish after integration. One can also see from these limits that to first order the Cooperon will give logarithmic terms in \(q\).

\begin{figure}
\centering
\includegraphics[width=\linewidth]{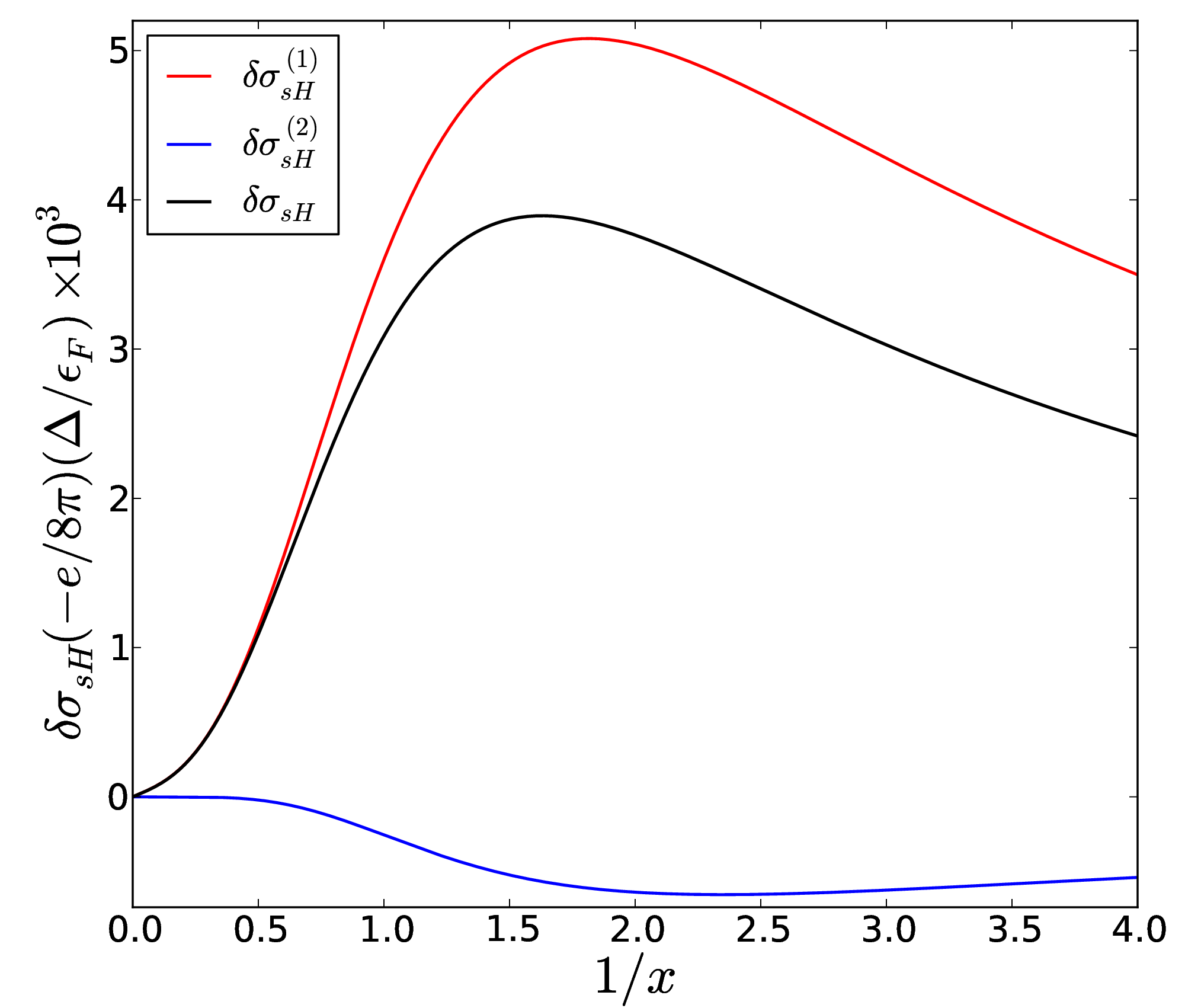}%
\caption{\label{f-Qcorrections}(color online) Quantum corrections to the spin Hall conductivity, for the first, second and third  (identical) Hikami boxes, and the total corrections. In units of \((-e/8\pi)(\Delta/\epsilon_F)\) the quantum corrections are function of \(x=\Delta \tau\) only.}
\end{figure}

Details on the computation of Eqs.~(\ref{eq.hikami1})-(\ref{eq.hikami3}), are given in the Appendix. We find that the quantum corrections contribute positively to the spin Hall conductivity (effect usually referred to as weak anti-localization). The corrections of all three boxes are proportional to the spin-orbit splitting energy \(\Delta/\epsilon_F\). We present in Fig.~\ref{f-Qcorrections} the plot of Eqs.~(\ref{eq.s1}) and (\ref{eq.s2}), that give the Hikami boxes corrections to the spin conductivity. The second and third boxes, that contribute equally to the corrections, are about one order of magnitude smaller than the corrections of the first box, and their contribution is negative. For large values of the parameter \(x\), [see Eqs.~(\ref{eq.s1}-\ref{eq.s2})],
\begin{align}
\delta \sigma^{(1)} _{sH} &= - \frac{-1}{8 \pi} \, \frac{\Delta}{\epsilon_F}\,\frac{ \left(8 \log \left(\frac{6}{5}\right)-7 \log \left(\frac{4}{3}\right)\right)}{784 \, x}\\
\delta \sigma^{(2)}_{sH}  &= \frac{-1}{8\pi} \, \frac{\Delta}{\epsilon_F} \,\frac{ \log \left(\frac{250}{243}\right)}{3528 \, x}\,,
\end{align}
the corrections tend to zero as \(\sim 1/x\). They also vanish for small values of \(x\),
\begin{equation}
\delta \sigma^{(1)} \sim \frac{-1}{8\pi} \frac{\Delta}{\epsilon_F} \frac{x}{64}, \quad
\delta \sigma^{(2)} \sim \frac{-1}{8\pi} \frac{\Delta}{\epsilon_F} \left(\frac{-x}{192}\right)\,.
\end{equation}
In the intermediate region, for \(x\) of order one, the quantum effects reach a maximum. It is interesting to note that the quantum corrections, up to a global factor \(\Delta/\epsilon_F\), are determined by a universal function, displayed in Fig.~\ref{f-Qcorrections}, that depends only on the ratio of the spin-orbit splitting to the disorder strength. The shape of this function, shows that increasing the disorder (with fixed spin-splitting), the quantum contribution to  the conductivity first increase, and then, after reaching a maximum, decrease.

It is worth noting that we calculate a spin conductivity and therefore the term (anti-) localization does not exactly have the same meaning as in the charge conductivity case. (Anti-) localization must be taken in the quantum spin transport sense and its meaning is more subtle than in the quantum charge transport picture.

%
%
\subsection{Spin conductivity}
\begin{figure}
\centering
\includegraphics[width=0.48\textwidth]{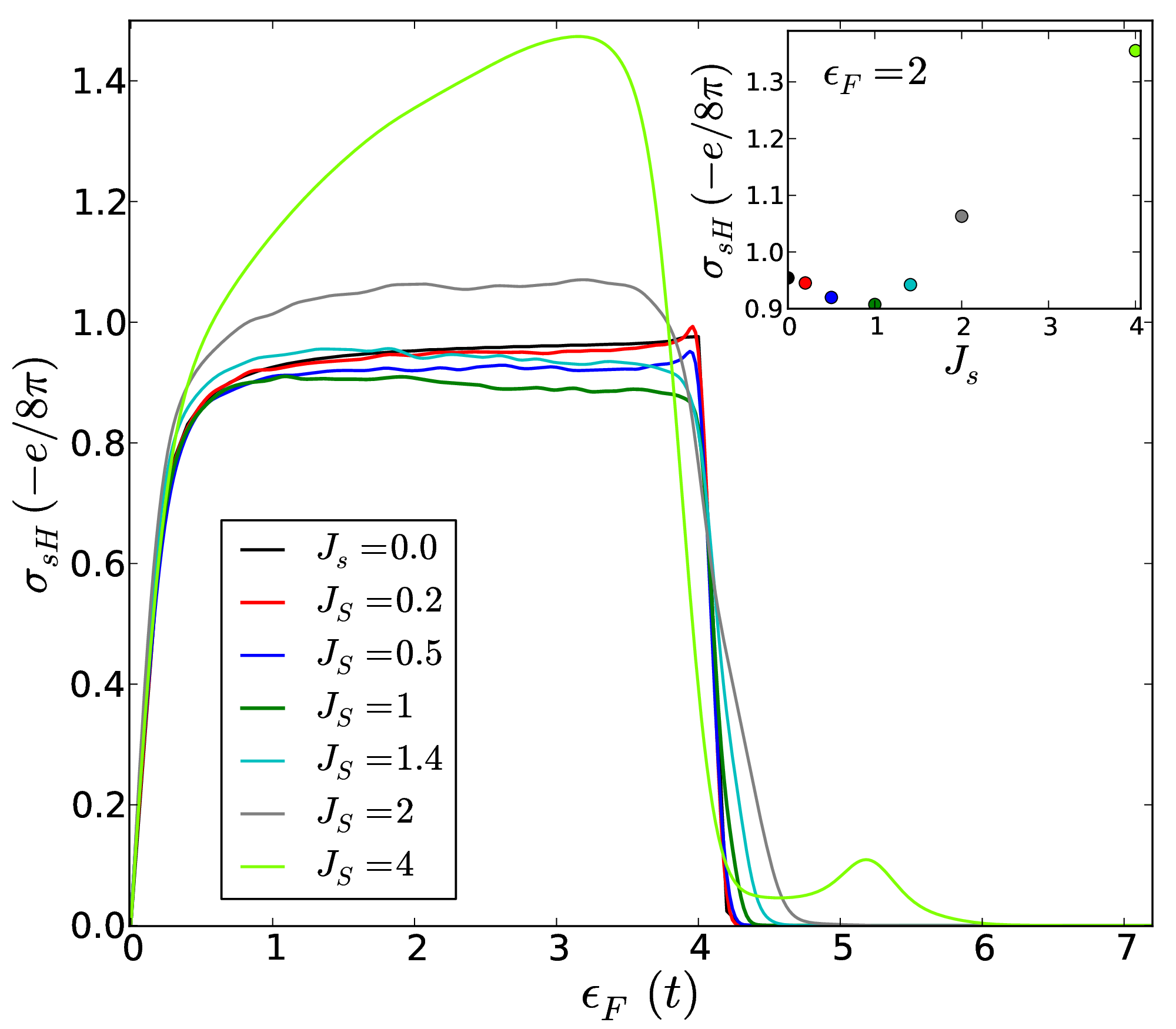}
\caption{(Color online) Static spin Hall conductivity as a function of the Fermi energy, for different values of the disorder strength. The inset show the variation of \(\sigma_{sH}\) at fixed \(\epsilon_F=2\); in the strong disorder regime it increases with the disorder strength.}
\label{f-s}
\end{figure}

Disorder has a very strong effect on the spin conductivity, especially for high values of the exchange energy, when separate impurity bands are present. For weak disorder we observe that the static spin conductivity slightly decreases with respect to the clean limit,\cite{van-den-Berg-2011fk} including a small modification due to quantum interference effects. However, at strong disorder the behavior of \(\sigma_{sH}\) qualitatively changes. In Fig.~\ref{f-s} we plot \(\sigma_{sH}\) as a function of the Fermi energy, for different values of the disorder strength, from the weak to the strong disorder limits. Increasing the disorder strength, for values of \(J_s<1\), \(\sigma_{sH}\) decreases, and for \(J_s>1\), it increases, showing that the localization transition at \(J_s\approx1\), modifies the behavior of the spin conductivity; this can be naturally attributed to the strong correlations of the carrier spin with the impurities, as discussed in Sec.~\ref{S-wave}, resulting from the fractal geometry of the quantum states and their localization. 

\begin{figure}
\centering
\includegraphics[width=0.48\textwidth]{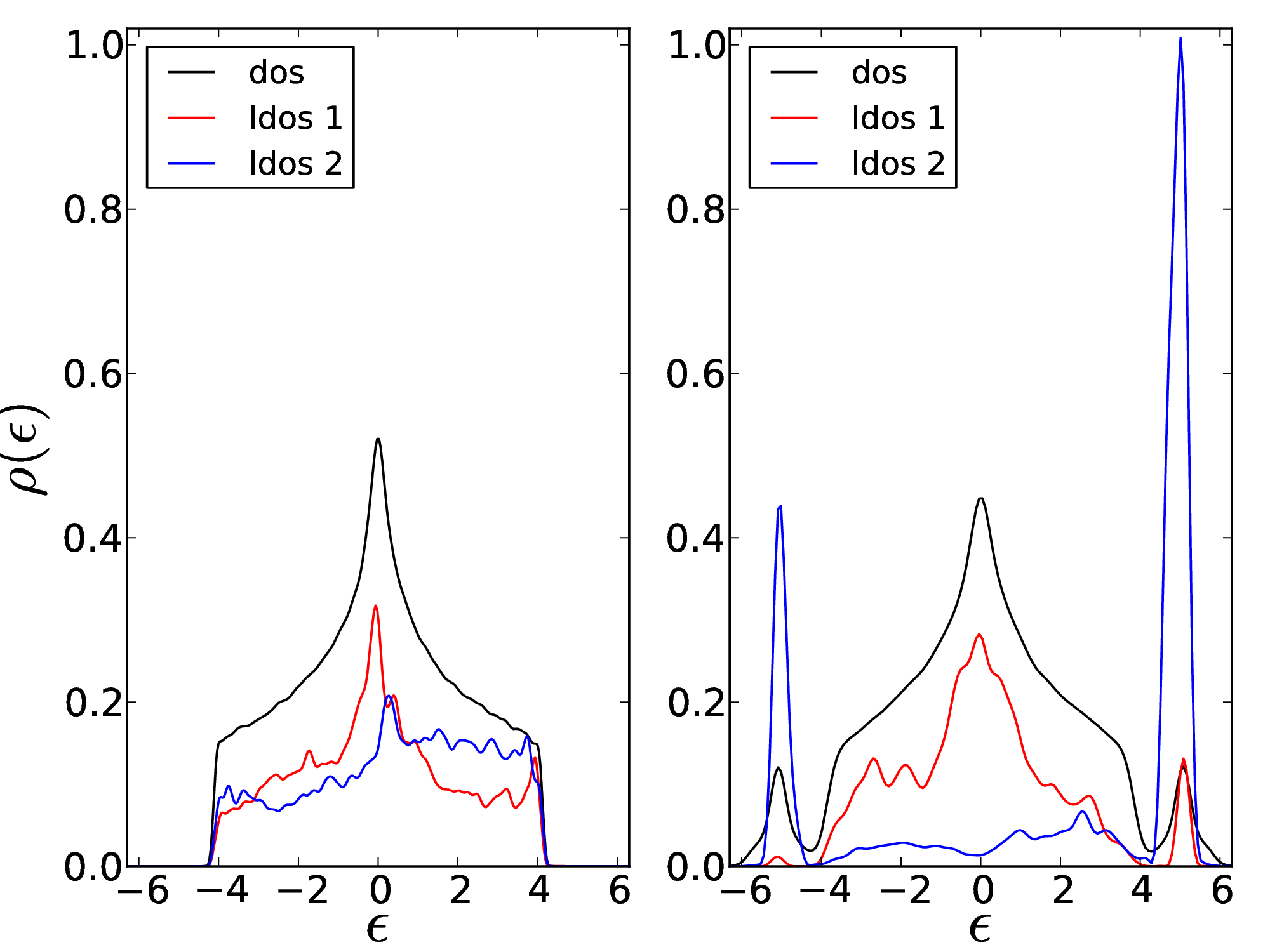}
\caption{(Color online) Local and total density of states for (left) \(J_s=1\), at the localization transition, and (right) \(J_s=4\), in the presence of impurity bands. The (unnormalized) local density of states is represented for two arbitrary sites in the lattice, occupied by impurities.}
\label{f-ib}
\end{figure}

If disorder is strong enough, the total density of states develops impurity bands; in this case, a striking intermittency in space and energy of the local density of states arises (Fig.~\ref{f-ib}). We observe that, near the localization transition, the energy distribution of states differs between sites, 
depending strongly on whether they are occupied or not by an impurity. For a given impurity, electronic states display a large distribution of energies. However, we observe important variations from site to site, showing a pronounced bias between positive and negative energy states. This systematic asymmetry is related to the spin dependence of the localized states. For strong disorder, there are sites for which the quantum states have energies almost entirely concentrated in the impurity bands, these states are strongly spin polarized.

The striking enhancement of the spin conductivity, starting at the localization transition, can be explained in terms of the strong spatial spin fluctuations at the impurity sites. Indeed, the spin current results from the imbalance between counter propagating spin-up and spin-down electrons; the presence of a magnetic impurity locally produces a marked bias in the spin orientation, due to the electron wave function localization: only states with the appropriate spin polarization are allowed. As we have shown before, quantum corrections, proportional to the spin-orbit splitting, tend to increase the spin conductivity; yet, in the strong disorder regime, the spin splitting is notably strengthened as a consequence of the localization of states around the impurities. The result is that, in the presence of an electric field, the usual torque mechanism that drives opposite spins to drift in opposite directions, transversally to the electric current,\cite{Sinova-2004fb} is greatly reinforced by the strong local spin splitting.

Finally, we find that the spin conductivity fluctuations increase with the disorder in the extended state regime, as already noted in Ref.~\onlinecite{van-den-Berg-2011fk}, but decrease above the localization transition. Concomitantly, we can follow the spin current associated with the wave packet, defined by \(J_x^z(t)=\langle\psi(t)|j_x^z|\psi(t)\rangle \),
\[
\bm J^z(t) = \frac{1}{2}\mathrm{Im} \, \int d\bm x \left[
	\psi_\uparrow^*\nabla\psi_\uparrow -
	\psi_\downarrow^*\nabla\psi_\downarrow
	\right]\,.
\]
This current vanishes in mean, but undergoes time fluctuations with amplitudes steadily increasing as a function of the disorder intensity, roughly linearly with \(J_s\) in the strong disorder range of parameters (at fixed concentration). This behavior is in accordance with the mechanism described above of the spin conductivity increase, as it is intimately related to the  spin current-charge current correlations.

%
%
\section{Discussion and conclusion}
\label{S-concl}
We considered a two-dimensional electron gas, with spin-orbit coupling and magnetic impurities, to study the spin Hall conductivity. We first investigated, using numerical methods to explore the whole range of parameters, the dynamical and spectral properties of the quantum states, as a function of the disorder strength. We then studied the corrections to the spin conductivity due to quantum interference in the weak disorder regime, using the linear response theory. Finally, we computed the spin conductivity in the strong disorder regime.

The evolution of a wave packet shows a rich dynamical behavior, with a disorder dependent spreading exponent, ranging from almost ballistic to slightly subdiffusive motion. This seems to be controlled by the modification introduced by the spin-orbit interaction, to the simple free motion. It is at variance with the usual diffusion observed in a random magnetic field, or even in disordered symplectic systems, and reminiscent to the behavior of quasi periodic systems. In addition, the geometry of the wave packet, as characterized by the correlation dimension, is fractal over the entire range of disorder strengths. In fact, we find that for a critical value of the exchange constant (at fixed impurity concentration), the system undergoes a localization transition. We observe that at this point the wave packet width and the correlation dimension are compatible with the characteristic diffusion exponents of a two-dimensional random walk. Above this point the dynamical inverse participation ratio remains finite at long times, indicating the spatial localization of the quantum states. This is confirmed by the spectral properties as displayed by the local density of states. In the localized regime, a strong intermittence arises (arithmetic and geometric means do not coincide), characterized be the appearance of strong spin correlations between carriers and impurities. 

It is natural to think that the interference effects observed in the phenomenology of the wave packet spreading, should modify the spin transport. Indeed, the analytical computation of the quantum corrections using the Kubo linear response theory, shows that these effects contribute to increase the spin conductivity, and that they are proportional to the spin-orbit splitting energy. Yet, the effect of the localization on the spin conductivity is rather surprising. We found that the spin conductivity actually pass a minimum at the transition, and strongly increases with disorder in the localized regime. The sticking of the electronic energy states to the impurities location, and the high spin polarization, contribute to reinforce the spin splitting and enhance the spin density fluctuations that, when an electric field is applied, will drift in opposite directions to create a strong spin current.

It would be interesting to generalize the present one particle model in a random potential to take into account the electron mediated interactions between magnetic impurities. This would allow us to investigate the interplay of Anderson localization and ferromagnetic transitions, and their influence on the spin transport, as revealed for instance in recent experiments.\cite{Yamada-2011fk}

%
%
\begin{acknowledgments}
We are grateful to R. Hayn and T. Martin for useful discussions. We thank X. Leoncini for his interest in this work.
\end{acknowledgments}

\appendix

\section{Cooperon}
The total \(\bm{q}\)-dependent Cooperon can be written in the form
\begin{equation}
\mathcal{C}(\bm{q}) = \begin{pmatrix}
\mathcal{C}^{\uparrow \uparrow}_{\uparrow \uparrow}&0&0&\mathcal{C}^{\uparrow \downarrow}_{\uparrow \downarrow}\\
0&\mathcal{C}^{\uparrow \uparrow}_{\downarrow \downarrow}&\mathcal{C}^{\uparrow \downarrow}_{\downarrow \uparrow}&0\\
0&\mathcal{C}^{\downarrow \uparrow}_{\uparrow \downarrow}&\mathcal{C}^{\downarrow \downarrow}_{\uparrow \uparrow}&0\\
\mathcal{C}^{\downarrow \uparrow}_{\downarrow \uparrow}&0&0&\mathcal{C}^{\downarrow \downarrow}_{\downarrow \downarrow}
\end{pmatrix}\,,
\label{eq.coopq}
\end{equation}
with
\begin{widetext}
\begin{align}
 \mathcal{C}_{\uparrow \uparrow}^{\uparrow \uparrow} &=\mathcal{C}^{\downarrow \downarrow}_{\downarrow \downarrow} = \frac{16 \left(x^2+1\right)^3 \left(\left(x^6+3 x^4+2\right) q^2 l^2+2 \left(x^2+1\right)^2 \left(5
   x^2+4\right)\right)}{\left(\left(x^6+3 x^4-6 x^2+4\right) q^2 l^2+4 \left(x^2+1\right)^2 \left(5
   x^2+4\right)\right) \left(\left(3 x^6+9 x^4+6 x^2+4\right) q^2 l^2+4 \left(x^2+1\right)^2 \left(5
   x^2+4\right)\right)} \nonumber \\
 \mathcal{C}_{\uparrow \downarrow}^{\uparrow \downarrow} &=\mathcal{C}^{\downarrow \uparrow}_{\downarrow \uparrow} = -\frac{8\, e^{-2 i \chi } q^2 l^2 x^2 \left(x^2+1\right)^3 \left(x^4+3
   x^2+6\right)}{\left(\left(x^6+3 x^4-6 x^2+4\right) q^2 l^2+4 \left(x^2+1\right)^2 \left(5
   x^2+4\right)\right) \left(\left(3 x^6+9 x^4+6 x^2+4\right) q^2 l^2 + 4 \left(x^2+1\right)^2 \left(5
   x^2+4\right)\right)} \nonumber \\
 \mathcal{C}^{\uparrow \uparrow}_{\downarrow \downarrow} &=\mathcal{C}^{\downarrow \downarrow}_{\uparrow \uparrow} = -\frac{\left(x^6+3 x^4+2\right) q^2 l^2+2 \left(x^3+x\right)^2}{(q^2 l^2- 4) \left(q^2 l^2 \left(3
   x^2-1\right)-2 \left(x^2+1\right)^2 \left(3 x^2+2\right)\right)} \nonumber \\
 \mathcal{C}^{\uparrow \downarrow}_{\downarrow \uparrow} &=\mathcal{C}^{\downarrow \uparrow}_{\uparrow \downarrow} = \frac{\left(10 x^4+28 x^2-q^2 l^2
   \left(x^4+3 x^2+6\right)+26\right) x^2+8}{(q^2 l^2 -4) \left(q^2 l^2 \left(3 x^2-1\right)-2
   \left(x^2+1\right)^2 \left(3 x^2+2\right)\right)} \nonumber
\end{align}
where \(\chi\) is the angle between the \(\bm{k}\) and \(\bm{q}\) and \(l=\tau p_F\) is the mean free path length. Because the interaction is isotropic on average, \(\mathcal{C}_{\uparrow \downarrow}^{\uparrow \downarrow}=0\) after integration over \(\chi\). The \(\bm{q}\)-integrated Cooperon is written
\begin{align}
\mathcal{C} = \frac{(1+x^2)^2}{8 \pi^2 N_F l^2 \tau} \left[ \frac{\log \left(\frac{(8 + 15 x^2 + 24 x^4 + 9 x^6)}{(1+x^2)^2 (8 + 12 x^2)}\right)}{2\left(3 x^4+8 x^2+13\right)} \right.
& \left[ \frac{(4+5x^2)}{x^2} \left( \sigma_x \otimes \sigma_x + \sigma_y \otimes \sigma_y \right) + \left(\mathbb{I} - \sigma_z \otimes \sigma_z \right) \right] \nonumber \\ 
&\qquad+ \left.\frac{\left(x^2+1\right) \log \left(\frac{(2 + x^2) (3 + 5 x^2 + 6 x^4)}{(1 + x^2)^2 (4 + 5 x^2)}\right)}{\left(x^6+3 x^4+2\right)} \left(\mathbb{I} + \sigma_z \otimes \sigma_z \right) \right] 
\label{eq.coop}
\end{align}
Using the \(\bm{q}\)-integrated Cooperon for the conductivity calculation and nondimensional expressions for the Green functions (Eq.~(\ref{eq.Green2}))
\begin{align}
\delta \sigma^{(1)}_{sH} &= -\int \frac{d \bm{k}}{(2\pi)^2} (G^A(\bm{k}, \epsilon_F) J_y^z (\bm{k}) G^R(\bm{k},\epsilon_F)) \otimes (G^R(-\bm{k}, \epsilon_F) J_x(-\bm{k}) G^A(-\bm{k},\epsilon_F) ) \, \mathcal{C} \\
 &=  -2 \pi N_F (2\tau)^3 \int \frac{d\xi}{2\pi} \int \frac{d\varphi}{2\pi} \left( G^A(\xi,\varphi) J_y^z G^R(\xi,\varphi)\right) \otimes \left( G^R(\xi,\varphi-\pi) J_x G^A(\xi,\varphi-\pi)\right) \\
\delta \sigma^{(2)}_{sH} &= - \int_{\Omega} \int_{\bm{k}} \int_{\bm{k'}}  (G^A(\bm{k})  J_y^z(\bm{k}) G^R(\bm{k}) V (\theta, \phi) G^R(\bm{k'})) \nonumber \\
&\qquad \otimes (G^R (\bm{q} -\bm{k'},) J_x(\bm{q}-\bm{k'}) G^A(\bm{q} -\bm{k'}) V (\theta, \phi) G^A(\bm{q} -\bm{k}) )\, \mathcal{C} \,, \nonumber \\
&= - \,(2\pi N_F)^2\,(2\tau)^6  \int_{\Omega} \int \frac{d \xi}{(2\pi)(2\tau)} \int \frac{d \xi'}{(2\pi)(2\tau)}  \int  \frac{d\varphi}{2\pi}  \int  \frac{d\varphi '}{2\pi} \nonumber \\
& \qquad (G^A(\xi,\varphi) \, V(\theta,\phi)\, G^A(\xi',\varphi ')\, J_y^z(k_F) \, G^R(\xi',\varphi '))\nonumber \\
&\qquad (G^R (\xi',\varphi ' -\pi)\,  V(\theta,\phi)\, G^R(\xi,\varphi-\pi)\, J_x(k_F) \,G^A(\xi,\varphi-\pi))\, \mathcal{C}\,,
\end{align}
The expression of the third diagram has the same structure as that of the second diagram. The first quantum corrections are written
\begin{align}
\delta\sigma^{(1)} _{sH} =& \frac{-1}{8\pi}\frac{\Delta}{\epsilon_F} \frac{\left(x^2+2\right)}{16 \,  x \left(x^2+1\right) \left(7 x^2+8\right)^2}
    \left[ \frac{8 \left(x^2+1\right) \left(x^4+x^2-4\right) x^4}{\left(2 + 3 x^4 + x^6 \right)} \log \left(\frac{\left(x^2+2\right) \left(6 x^4+5 x^2+3\right)}{\left(x^2+1\right)^2 \left(5 x^2+4\right)}\right) \right. \nonumber \\
    & \qquad + \frac{ \left(11 x^8+63 x^6+88 x^4+80 x^2+32\right)}{\left(13 + 8 x^2 + 3 x^4\right)} \log \left(\frac{9 x^6+24 x^4+15 x^2+8}{12 x^2+8}\right) \nonumber \\
    & \qquad - \frac{2 \left(5 x^2+4\right) \left(x^6+5 x^4+4 x^2+4\right)}{\left(13 + 8 x^2 + 3 x^4\right)} \log \left(\frac{12 x^2+8}{9 x^6+24 x^4+15 x^2+8}\right) \nonumber \\
   & \qquad -\left. \frac{ 2 \left(21 x^8+121 x^6+168 x^4+152 x^2+64\right)}{\left(13 + 8 x^2 + 3 x^4\right)} \log \left(x^2+1\right) \right]
\label{eq.s1}
\end{align}
The second and third Hikami boxes give an equal contribution
\begin{align}
\delta \sigma^{(2)}_{sH}& = \frac{-1}{8\pi}\frac{\Delta}{\epsilon_F}\frac{\left(x^2+2\right) }{24 \,x \left(x^2+1\right) \left(7 x^2+8\right)^2}  \left[ \frac{2 \left(x^8+x^6+9 x^4+12 x^2+8\right)}{\left(3 x^4+8 x^2+13\right)} \log \left(\frac{\left(1+x^2\right)^2 \left(12 x^2+8\right)}{9 x^6+24 x^4+15 x^2+8 }\right)  \right.\nonumber \\
   & \qquad - \left. \frac{\left(3 x^{10}+14 x^8+32 x^6+22 x^4-19 x^2-52\right) x^4}{ \left(3
   x^4+8 x^2+13\right) \left(x^6+3 x^4+2\right)} \log \left(\frac{\left(x^2+2\right) \left(6 x^4+5 x^2+3\right)}{\left(x^2+1\right)^2
   \left(5 x^2+4\right)}\right)\right]
\label{eq.s2}\,.
\end{align}
All corrections scale as \(\Delta/\epsilon_F\).

\end{widetext}

%
%

%

\end{document}